\documentclass{aastex}
\usepackage{emulateapj5}
\usepackage{apjfonts}

\def\wig#1{\mathrel{\hbox{\hbox to 0pt{%
          \lower.5ex\hbox{$\sim$}\hss}\raise.4ex\hbox{$#1$}}}}

\shorttitle{Three-Dimensional Transmission Spectra}
\shortauthors{Fortney, et al.}

\newcommand{\rj}{$R_{\mathrm{J}}$}
\newcommand{\me}{$M_{\oplus}$}

\newcommand{\rp}{$R_{\mathrm{p}}$}
\newcommand{\rs}{$R_{\mathrm{S}}$}

\newcommand{\hd}{HD 209458b} 

\newcommand{\he}{HD 189733b}

\newcommand{\te}{$T_{\rm eff}$}

\newcommand{\cp}{\citep}
\newcommand{\ct}{\citet}

\begin{document}

\title{Transmission Spectra of Three-Dimensional Hot Jupiter Model Atmospheres}

\author{J. J. Fortney\altaffilmark{1}, M. Shabram\altaffilmark{1}, A. P. Showman\altaffilmark{2}, Y. Lian\altaffilmark{2}, R. S. Freedman\altaffilmark{3}$^{,}$\altaffilmark{4}, M. S. Marley\altaffilmark{3}, N. K. Lewis\altaffilmark{2}} 

\altaffiltext{1}{Department of Astronomy and Astrophysics, University of California, Santa Cruz, CA 95064; jfortney@ucolick.org}
\altaffiltext{2}{Lunar and Planetary Laboratory, 1629 E. University Blvd., University of Arizona, Tucson, AZ}
\altaffiltext{3}{Space Science and Astrobiology Division, NASA Ames Research Center, Mail Stop 245-3, Moffett Field, CA 94035}
\altaffiltext{4}{SETI Institute, 515 N. Whisman Road, Mountain View, CA 94043}

\begin{abstract} 
We compute models of the transmission spectra of planets \hd, \he, and generic hot Jupiters.  We examine the effects of temperature, surface gravity, and metallicity for the generic planets as a guide to understanding transmission spectra in general.  We find that carbon dioxide absorption at 4.4 and 15 $\mu$m is prominent at high metallicity, and is a clear metallicity indicator.  For \hd\ and \he, we compute spectra for both one-dimensional and three-dimensional model atmospheres and examine the differences between them.  The differences are usually small, but can be large if atmospheric temperatures are near important chemical abundance boundaries.  The calculations for the 3D atmospheres, and their comparison with data, serve as constraints on these dynamical models that complement the secondary eclipse and light curve data sets.  For \hd, even if TiO and VO gases are abundant on the day side, their abundances can be considerably reduced on the cooler planetary limb.  However, given the predicted limb temperatures and TiO abundances, the model's optical opacity is too high.  For \he\ we find a good match with some infrared data sets and constrain the altitude of a postulated haze layer.  For this planet, substantial differences can exist between the transmission spectra of the leading and trailing hemispheres, which is an excellent probe of carbon chemistry.  In thermochemical equilibrium, the cooler leading hemisphere is methane-dominated, and the hotter trailing hemisphere is CO-dominated, but these differences may be eliminated by non-equilibrium chemistry due to vertical mixing.  It may be possible to constrain the carbon chemistry of this planet, and its spatial variation, with \emph{JWST}.

\end{abstract}


\section{Introduction}
Soon after the detection of the transits of planet \hd\ \cp{Charb00,Henry00}, atmosphere modelers realized that transits could provide a probe of the atmospheric composition of transiting planet atmospheres \cp{SS00,Brown01,Hubbard01}.  Stellar light, which passes through the thin upper atmosphere of a transiting planet, travels through a long path length in the planet's atmosphere, which allows atomic and molecular absorption features in the  atmosphere to be imprinted on top of the stellar spectrum.  This prediction was stunningly confirmed by observations of \hd\ by \ct{Charb02} with the STIS instrument aboard the \emph{Hubble Space Telescope}.

This first detection of an exoplanet's atmosphere began our current era of exoplanet characterization, which has seen our knowledge of these planets expand due to \emph{Hubble}, \emph{Spitzer}, \emph{MOST}, \emph{CoRoT}, \emph{Kepler}, and ground-based efforts.  The use of \emph{Hubble}'s STIS spectrograph as the premier instrument of transmission spectroscopy was cut short by the instrument's failure in 2004, soon after the detection of an evaporating exosphere around \hd\ \cp{Vidal03,Vidal04} and the full-optical transmission spectrum of \hd\ \cp{Knutson07a}.  The \ct{Knutson07a} data set was further analyzed by \ct{Barman07} and D.~Sing and collaborators \cp{Sing08a,Sing08b,Lecavelier08b,Desert08}.

For \hd, the optical transmission spectra data sets achieved so far has basically confirmed the predictions of the first models \cp{SS00,Brown01,Hubbard01}: absorption features, which can be thought of as an increase in the cross-sectional area or radius of the planet at a given wavelength, can be seen during the transit.  Prominent features include Rayleigh scattering off of H$_2$ molecules in the blue, absorption due to the neutral atomic Na doublet at 589 nm, and perhaps TiO/VO absorption \cp{Desert08} and H$_2$O absorption in the red \cp{Barman07}.  \ct{Zahnle09} have also pointed out that species such as HS may also contribute in the blue, along with the Rayleigh scattering contribution.

In the absence of this optical space spectrograph, other instruments aboard \emph{Hubble} and \emph{Spitzer} were used to probe the atmosphere of planet \he.  \emph{Hubble}'s NICMOS was used by \ct{Swain08} to obtain a near IR transmission spectrum, which detected CH$_4$ \cp[but see also][]{Sing09b}.  Using \emph{Hubble} ACS \ct{Pont08} detected a featureless, sloping optical spectrum, perhaps implying a Rayleigh-scattering haze layer, and \ct{Tinetti07} and \ct{Beaulieu08} used \emph{Spitzer}'s IRAC to find absorption variation in the mid IR that is consistent with absorption via water vapor \cp[however, see also][for a different analysis]{Desert09}.  For \hd\ at mid-infrared wavelengths, there is a 24 $\mu$m MIPS point \cp{Richardson06} and a four-band IRAC data set \cp{Beaulieu09}.  \hd, a pM class planet in the terminology of \ct{Fortney08a} and \he, pL class, remain by far the best characterized transiting planets, and this will likely remain the case for quite some time.

The characterization of these atmospheres has increased in sophistication, in particular with the complementary detection of secondary eclipses of the transiting planets, which yield the day-side planet-to-star flux ratios for these systems \cp[e.g.][]{Knutson08,Charb08,Machalek08}.  These kinds of observations are complementary to the transmission spectrum observations.  Due to the longer path length at slant viewing geometry, transmission spectra probe lower atmospheric pressures at a given wavelength \cp{Fortney05c}.  Perhaps more importantly, transmission spectra probe the terminator region of the planets, the great circle route dividing day from night.  The terminator region is obviously much less irradiated than the substellar point or the day-side mean, so we could expect temperature dependent and irradiation dependent chemical mixing ratios that are different on the terminator, compared to the substellar point, due to the lower incident fluxes and cooler temperatures.  Interestingly, \ct{Swain08} and \ct{Swain09} apparently find such a situation for \he, where on the day side, they find little to no carbon in CH$_4$, while it is detectable at the terminator.

More complex models of the atmospheric temperature structure, chemical mixing ratios, and the dynamics of these atmospheres have become necessary to understand these growing data sets.  Recent one-dimensional models of hot Jupiters, that aimed to predict the mean temperature structure and spectrum \cp{Burrows05b,Fortney05,Seager05}, and ignored dynamics, have yielded in some circumstances to multi-dimensional purely radiative-transfer models \cp{Barman05}, or models that also include parameterizations of dynamics \cp{Iro05,Burrows08}.

Our group has taken a different path, first in examining the spectral properties of three-dimensional dynamical models \cp{Fortney06b,Showman08}, and now in pioneering the coupling of non-gray radiative transfer to three-dimensional dynamics \cp{Showman09}.  While these 3D models are being tested by observations of secondary eclipses and orbital phase variations with \emph{Spitzer}, it is also possible to test their predictions for chemical mixing ratios and upper atmospheric temperatures via transmission spectra observations.  The main focus of this paper is to describe the results of the calculation of the transmission spectra of these three-dimensional models.

The paper is organized as follows.  In \S2 we describe our methods, while in \S3 we show results for the transmission spectrum of simple isothermal models to investigate the role of surface gravity and enhanced metallicity.  In \S4 we describe our three-dimensional atmosphere models, while in \S5 we review the transmission spectrum data for planet \he, and present our models, while \S6 is the corresponding section for planet \hd.  \S7 focuses on the observability of chemical abundance gradients in hot Jupiter atmospheres while \S8 gives our conclusions.

\section{Methods: Transmission Spectrum}
We have previously developed a code to compute the transmission spectrum of planetary atmospheres.  The first generation of the code, which used one-dimensional atmospheric pressure-temperature (\emph{P-T}) profiles, is described in \ct{Hubbard01}.  In addition to absorption by atomic and molecular species, \ct{Hubbard01} also explored the effects of refraction and a glow of photons around the planet's limb due to Rayleigh scattering.  These additional effects were found to be negligible for hot Jupiters, so we ignore them here.  \ct{Fortney03} investigated simple two-dimensional models of the atmosphere of \hd, which included one planet-wide \emph{P-T} profile, but 2D changes in the absorption cross-section of neutral Na, due to photoionization of Na as function of depth and angle from the substellar point.

One can imagine a straight path through the planet's atmosphere, parallel to the star-planet-observer axis, at an impact parameter $r$ from this axis.  The gaseous optical depth $\tau_{\rm G}$, starting above the terminator and moving outward in one direction along this path, can be calculated via the equation:
\begin{equation} \label{taum}
\tau_{\rm G}=\int_{r}^{\infty}\frac{r^{\prime}dr^{\prime}\sigma(r^{\prime}) n(r^{\prime})}
{{(r^{\prime2}-r^{2})}^{1/2}},
\end{equation}
where $r^{\prime}$ is the distance between the local location in the atmosphere and the planetary center, $n$ is the local number density of molecules in the atmosphere, and $\sigma$ is the wavelength-dependent cross-section per molecule.  We assume hydrostatic equilibrium with a gravitational acceleration that falls off with the inverse of the distance squared.\footnote{While the transmission spectrum calculation allows for a decreasing surface gravity with height, the  calculation the initial \emph{P-T} profiles does not.  The change in gravity with height is a minor effect, so this does not lead to a significant inconsistency.}  In practice, the model atmosphere is broken up into 1000 radius intervals, between the arbitrary base radius $r$ and the maximum radius $r + \Delta r$, where $\Delta r$ is 10-15$\times 10^3$ km.  The base radius, which is either at 10 or 100 bars, is adjusted to yield the best fit to the observations.  Along the light ray path through the atmosphere parallel to the star-planet-observer axis, the local atmospheric density and opacity are also sampled at 1000 points.  For absorption, the wavelength dependent cross-section is calculated based on the local chemical equilibrium mixing ratios, while the Rayleigh scattering cross-section is
\begin{equation}
\sigma_{\rm R}={{8 \pi^3 (2+\nu)^2 \nu^2} \over {3 \lambda^4 n^2}},
\end{equation}
where $\nu$ is the refractivity (refractive index minus 1) of the gas.
Since $\nu \ll 1$ and $\nu \propto n$, $\sigma_{\rm R}$ is
a function only of $\lambda$ and the gas composition.

Here we define the wavelength-dependent transit radius as the radius where the total slant optical depth reaches 0.56, following \ct{Lecavelier08b}.  We match the analytical results of these authors for the transit radius as a function of wavelength for a constant gravity, pure Rayleigh scattering atmosphere.

\begin{figure}
\epsscale{.75}
\plotone{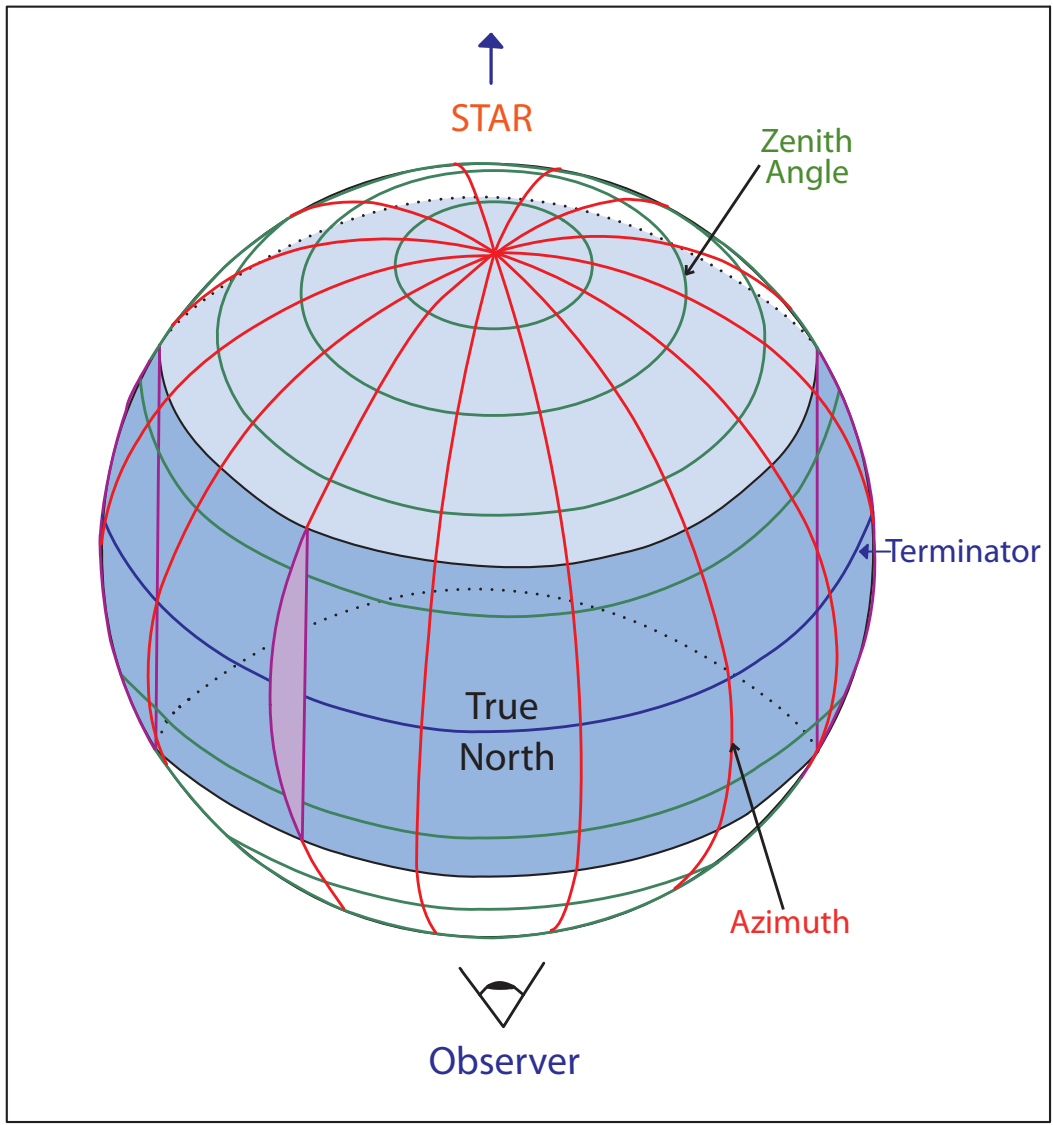}
\caption{Diagram of our zenith angle and azimuth grid, that aids in the computation of 3D transmission spectra.  The antistellar point faces the observer, while the substellar point face the star.  Azimuths are in red, while zenith angles are in green.  The terminator is shown in blue and the planet's true north pole is also labeled.  The blue band and purple slice are meant to schematically show the region of the atmosphere that is being probed during the transit.
\label{pic}}
\end{figure}

Now we have modified and generalized the code to compute transmission spectra for a three-dimensional temperature structure on an arbitrary latitude and longitude grid.  We define a grid based on the star-planet-Earth axis, rather than the planet's rotation axis.  The planet's substellar point is where the stellar \emph{zenith angle} is zero.  The zenith angle increases to 90 at the terminator and 180 and the antistellar point, which faces Earth during the transit.  We additionally define a \emph{azimuth} whose poles are the substellar and antistellar points.  See Figure \ref{pic}.   The \emph{true} north pole of the planet (defined by the rotation axis) runs through the azimuth=0 (prime) meridian.  Azimuth values increase towards the trailing hemisphere.  Using this new grid, we can first simply calculate the wavelength-dependent transmission of the planet at an azimuth, as a function of zenith angle.  In practice only zenith angles from $\sim$~70 to 110 degrees are sampled on the limb.  Once the planet radius as a function of azimuth is determined, we then average these radii around the azimuths of the planet, to find the mean planetary wavelength-dependent radius.

For any particular column of atmosphere, which is at a particular azimuth/zenith angle location, hydrostatic equilibrium is assumed, and we use the given local \emph{P-T} profile to interpolate in a pre-tabulated chemical equilibrium and opacity grid that extends out to 1 $\mu$bar.  The equilibrium chemistry mixing ratios \cp{Lodders02,Lodders06,Lodders09} are paired with the opacity database \cp{Freedman08} to generate pressure-, temperature-, and wavelength-dependent absorption cross-sections that are used for that particular column.  For a different column of atmosphere, with a different \emph{P-T} profile, local pressures and temperatures will yield different mixing ratios and wavelength-dependent cross-sections.

\section{Calculations for Simple Atmospheres} \label{simp}
\begin{figure}
\plotone{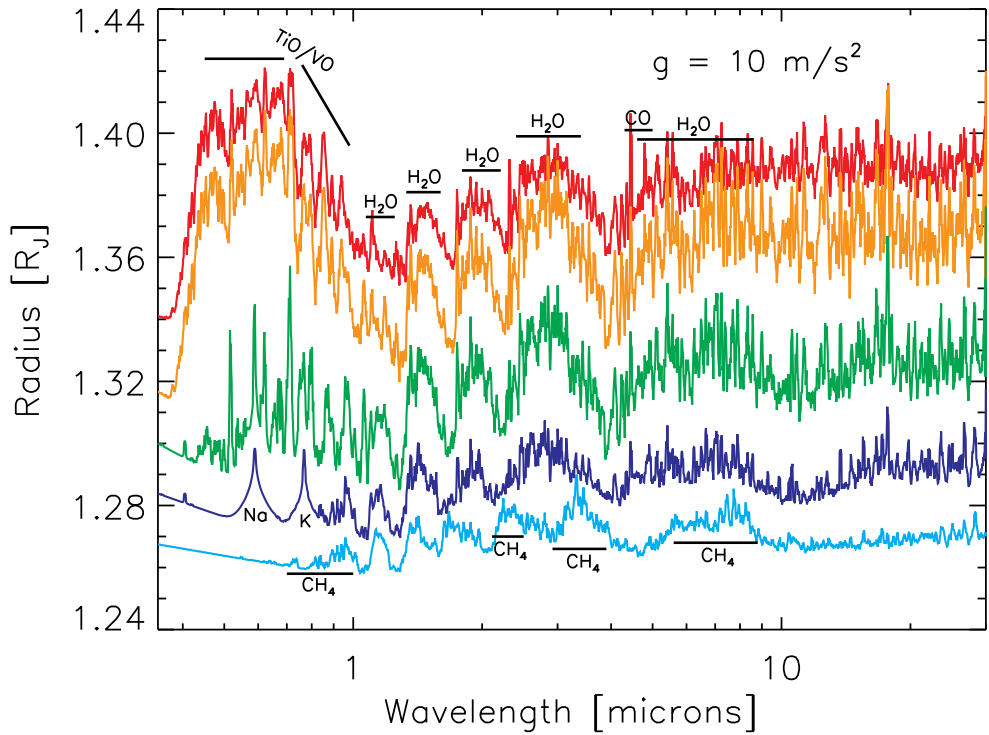}
\caption{Planet radius vs.~wavelength for isothermal model atmospheres with a gravitational acceleration of 10 m s$^{-2}$.  Models are $T$=2500, 2000, 1500, 1000, and 500 K, from top to bottom and assume chemical equilibrium.  Prominent absorption features, which appear as increases in the planetary radii, are labeled.
\label{g10}}
\end{figure}

\begin{figure}
\plotone{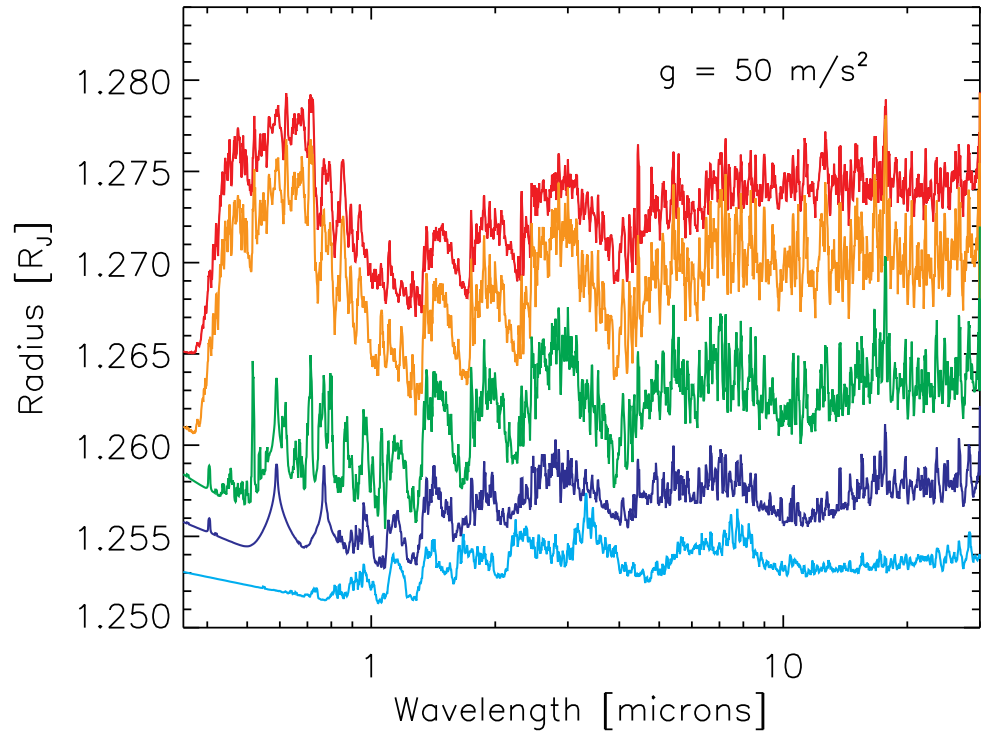}
\caption{Planet radius vs.~wavelength for isothermal model atmospheres with a gravitational acceleration of 50 m s$^{-2}$.  Models are $T$=2500, 2000, 1500, 1000, and 500 K, from top to bottom and assume chemical equilibrium.
\label{g50}}
\end{figure}

Our survey of the literature finds that there has been no systematic exploration of the change of transmission spectra with planetary \te\ for close-in planets.  Other authors have previously explored reflection and emission spectra \cp[e.g.][]{Marley99,Sudar00,Barman01,Sudar03,Marley07b}, but here we briefly examine transmission spectra for these atmospheres.  To maintain simplicity, we have investigated isothermal model atmospheres at 2500, 2000, 1500, 1000, and 500 K at two surface gravities, $g=10$ and 50 m s$^{-2}$.  The chemistry and opacity databases are the same as those used for the 3D model atmospheres.  The results of these calculations are shown in Figures \ref{g10} and \ref{g50}, for the low and high gravity cases, respectively.  All model planets have a base radius of 1.25 \rj\ at a pressure of 10 bars.

The striking impact of planetary surface gravity is clear from a comparison between these plots.  While the absorption features (where the radius is larger) are the same, the dynamic range is far larger in the lower gravity planets.  As previously discussed by other authors \cp[e.g.][]{SS00}, the region of the planetary atmosphere probed in transmission is an annulus with an area proportional to the atmospheric scale height.  Since lower gravity planets have larger scale heights, they are much better targets for transmission spectroscopy.

In these calculations we assume chemical equilibrium.  In the optical, this yields absorption bands of TiO and VO that are quite strong at 2000 and 2500 K, and weak at 1500 K.  Sodium and potassium doublets are clearly strong at 1000 and 1500 K.  In the 500 K model only bands of CH$_4$ and H$_2$O remain.  In the blue-optical, Rayleigh scattering can be seen as a monotonic increase in radius towards the blue, although in the hottest models TiO absorption swamps most of this effect until blue-ward of 400 nm.  The slope of this Rayleigh scattering region is indicative of the temperature, and becomes steeper (a more negative slope) as temperature increases \cp{Lecavelier08a}.  Also importantly the TiO/VO cross-sections fall off dramatically blue-ward of 0.4 $\mu$m.  The detection of a dramatic change in radius here would likely indicate TiO absorption.  In the infrared, bands of H$_2$O and CO are strong in the models from 1500-2500 K, while CH$_4$ replaces CO in the 500 and 1000 K models.  Figures \ref{g10} and \ref{g50} should serve as more accurate replacements for the transmission spectra of pM and pL class planets in Figure 11 of \ct{Fortney08a}.

\begin{figure}
\plotone{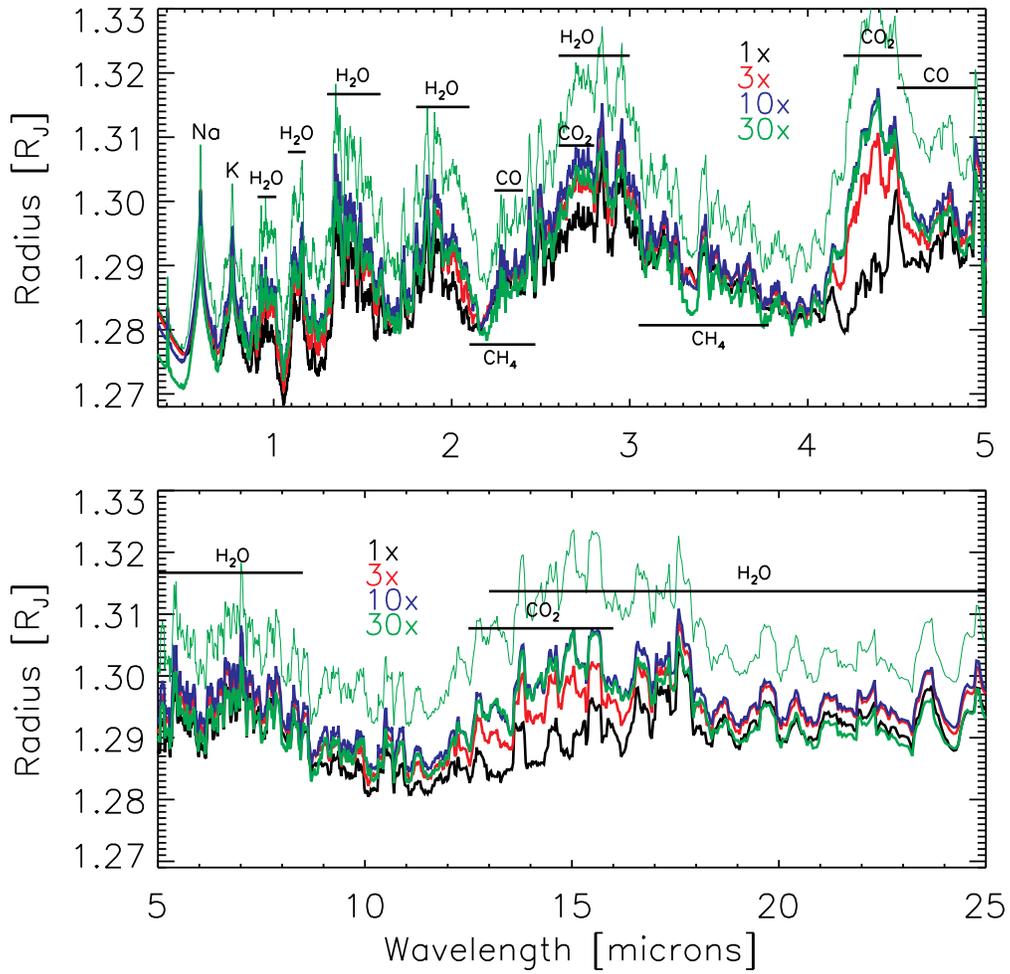}
\caption{Planet radius vs.~wavelength for the 1000 K isothermal model atmosphere at 4 values of metallicity.  For the thick lines, the correct mean molecular weight is calculated for each metallicity.  The thin green line is the 30$\times$ solar model, but with an unphysically low mean molecular weight (that of the 1$\times$ model) for comparison.
\label{metal}}
\end{figure}

We next turn our attention to the effects of atmospheric metallicity for H/He dominated atmospheres, as expected for close-in Jupiter- and Neptune-class planets.  \ct{MillerRicci09a} have previously investigated the transmission spectra of 1-10 \me\ planets with hydrogen-rich and hydrogen-poor atmospheres, and found that very large mean molecular weight (MMW) of hydrogen-poor atmospheres makes them extremely difficult targets for transmission spectroscopy, due to the much reduced scale height.  In \mbox{Figure \ref{metal}} we investigate hydrogen-rich 1000 K atmospheres from 1$\times$ to 30$\times$ solar metallicity.  We find that while the MMW climbs from 2.35 to 2.97 in this case, the transmission spectra signatures are, to first order, not hampered.

\mbox{Figure \ref{metal}} shows a 1$\times$ model in black and 30$\times$ model, with this same MMW, in thin green.  Radius values at all wavelengths increase due to the higher opacity everywhere.  The thick green curve, with the correct MMW, brings these radii down to values similar to the 1$\times$ model.  Of interest at 1000 K is that the atmosphere becomes much more CO and CO$_2$ dominated as the metallicity increases, since CO becomes more abundant linearly, and CO$_2$ quadratically, with metallicity \cp[e.g.][]{Lodders02,Zahnle09}.  The strong CO$_2$ bands at 4.4 and 15 $\mu$m are excellent and clear probes of atmospheric metallicity.  In the future, achieving spectra between 4 and 5 $\mu$m will particularly useful, as CO is also very strong from 4.5 to 5 $\mu$m.  The IRAC 4.5 $\mu$m band is so wide that it encompasses the CO and CO$_2$ absorption features.

\section{Methods: Atmosphere Models}
We have run three-dimensional numerical simulations using the
Substellar and Planetary Atmospheric Radiation and Circulation
(SPARC) model \cp{Showman09}, which couples the MITgcm \cp{Adcroft04} to the plane-parallel, multi-stream
radiative-transfer code of \ct{MM99}.  The MITgcm
is a state-of-the-art atmospheric and oceanic circulation model that
solves the global primitive equations, which are the standard equations
for the large-scale dynamics in planetary atmospheres,
in spherical geometry using pressure as a vertical coordinate.
The primitive equations are generally appropriate when the atmosphere
is stably stratified and when the horizontal/vertical aspect ratio of
the atmospheric circulation is large, as is expected to be true on many
hot Jupiters. The equations are discretized using finite-volume methods,
which cast the equations into a flux form that has good conservation
properties, and adopt a ``cubed sphere'' grid rather than the more
commonly used longitude-latitude grid, which allows longer time-steps and
increases the accuracy near the poles.  After completion of a simulation, the output can be interpolated onto a regular latitude/longitude grid.  See \ct{Showman09} for
additional details.

Many previous atmospheric circulation models have forced the
dynamics using highly idealized prescriptions for the thermodynamic
heating/cooling rate, such as a Newtonian relaxation scheme
\cp{Showman02,CS05,CS06,Showman08,Menou08,Rauscher09} or flux-limited diffusion
\cp{Dobbs08}.  In contrast, the SPARC/MITgcm model instead
determines the thermodynamic heating/cooling rate using the state-of-the-art
non-gray radiative transfer scheme of \ct{MM99}.  The
three-dimensional temperature structure can be viewed as numerous
one-dimensional vertical columns (one per grid-box on the horizontal grid).
At each model time-step, the model calculates the upward and downward
intensities versus wavelength for each such vertical column.  To
calculate the thermodynamic heating rate over the full 3D grid, the
model then sums the intensities over wavelength and takes the negative of the vertical divergence
of the resulting net flux; this heating/cooling rate (which generally
gives heating on the dayside and cooling on the nightside) is then used
in the thermodynamic energy equation to drive the dynamics.

The planet's thermal radiative transfer is calculated for wavelengths from 0.26 to 300 $\mu$m, with incident stellar flux treated from 0.26 to 6 $\mu$m,
using a correlated-k method to treat the opacities \cp{Goody89}.  A full description of the opacity database is given in \ct{Freedman08}.  The
three-dimensional simulations described here, as in Showman et al. (2009),
use 30 wavelength bins (see Showman et al.~2009 for more details).  For a given
metallicity,
the chemical mixing ratios and opacities are calculated assuming local chemical equilibrium
as a function of local temperature and pressure, accounting properly
for depletion of gas-phase species via condensation, follow K.~Lodders and collaborators \cp{Lodders99,Lodders02,Lodders02b,Visscher06,Lodders09}.  Note, however,
that we include only gas-phase opacity; any opacity from clouds/hazes
is neglected.

Local chemical equilibrium is likely a simplification for these highly irradiated, dynamic atmospheres.  For instance \ct{CS06} have previously shown that carbon chemistry may be out of equilibrium, with CO favored at the expense of CH$_4$, as is known for brown dwarfs of similar \te\ \cp{Saumon03,Saumon06,Hubeny07,Geballe09}.  This process is also well-studied in the solar system \cp[e.g.][]{Prinn77,Yung88,Bezard02}.  Alternatively, chemical mixing ratios may be modified by photochemistry \cp[e.g.][]{Liang04,Zahnle09}.  The importance of photochemistry at the particular location that we are interested in here, the terminator, is not clear.  Given these caveats, for \he\ in particular we will investigate the transmission spectra assuming both chemical equilibrium and a fixed CO/CH$_4$ ratio, following \ct{CS06}.  Given the importance of CO$_2$ at high metallicity, future work on nonequilibrium chemistry should include the effect of this molecule as well.

Here, we analyze SPARC/MITgcm simulations of HD 189733b and HD 209458b
that adopt solar metallicity and use a horizontal resolution of C32,
approximately equivalent to a standard global grid of 128 longitudes $\times$ 64 latitudes.  (Here is interpolated onto 64 zenith angles and 128 azimuths.)  The simulations use 53  vertical layers, with a base at 200 bar and a top at 2 $\mu$bar (see Showman et al.~2009).  Our HD 189733b case excludes TiO and VO
from the opacity calculation, while the HD 209458b case includes both
these species due to their possible relevance in producing the stratosphere
inferred on that planet \cp{Knutson08,Burrows07c,Fortney08a,Showman09,Spiegel09}.  Our HD 209458b simulation
is identical to that in Showman et al.~(2009) (see their Figs. 16-21);
our HD 189733b simulations is analogous to their nominal HD 189733b
case (see Figs. 4-10 in Showman et al.~2009) except that it includes more
vertical layers and extends to 2 $\mu$bar rather than 0.2 mbar.  Both simulations assume tidal locking of the planets---i.e., a permanent day and night hemisphere.

In \ct{Showman09} we considered the thermal emission of these planets as a function of orbital phase, post-processing the \emph{P-T} grids to compute spectra as a function of orbital phase, using the methods outlined in \ct{Fortney06b}.  These spectra can also be integrated over \emph{Spitzer} bandpasses to compute thermal emission light curves.  These model spectra and lightcurves can then be directly compared to observed secondary eclipse photometry, spectra, and lightcurves.  The main goals of this work is to further constrain the 3D model atmospheres in a region of the atmosphere that is distinct from that probed at secondary eclipse, as well as to point to future observations that may help us better understand these planetary atmospheres.

\section{\he} \label{sec189}
Since its detection by \ct{Bouchy05}, planet \he\ has lept to the forefront of exoplanetary characterization.  It features a large planet-to-star radius ratio and bright parent star, which enables detailed study of its atmosphere via observations of both its emission and transmission spectrum.  For the transmission spectrum, optical data have been published by \ct{Pont08} and \ct{Redfield08}, near infrared-data by \ct{Swain08} and \ct{Sing09b}, and 3.6 and 5.8 $\mu$m IRAC mid-infrared data by \ct{Beaulieu08} and \ct{Ehrenreich07}---this \emph{Spitzer} data was then reduced again by \ct{Desert09}, who also added new 4.5 and 8.0 $\mu$m points.

Soon after the discovery of \he, \ct{Fortney06} computed the first models of its atmosphere.  They found that the planet was considerably cooler than that of \hd, and was likely more similar to TrES-1, which had just been observed in secondary eclipse \cp{Charb05}.  In particular, \ct{Fortney06} noted the strong methane absorption features in their 1D model, which was meant to represent a planet-wide average \emph{P-T} profile.  This prediction was confirmed with the \ct{Swain08} detection of methane in the planet's transmission spectrum.

\begin{figure}
\plotone{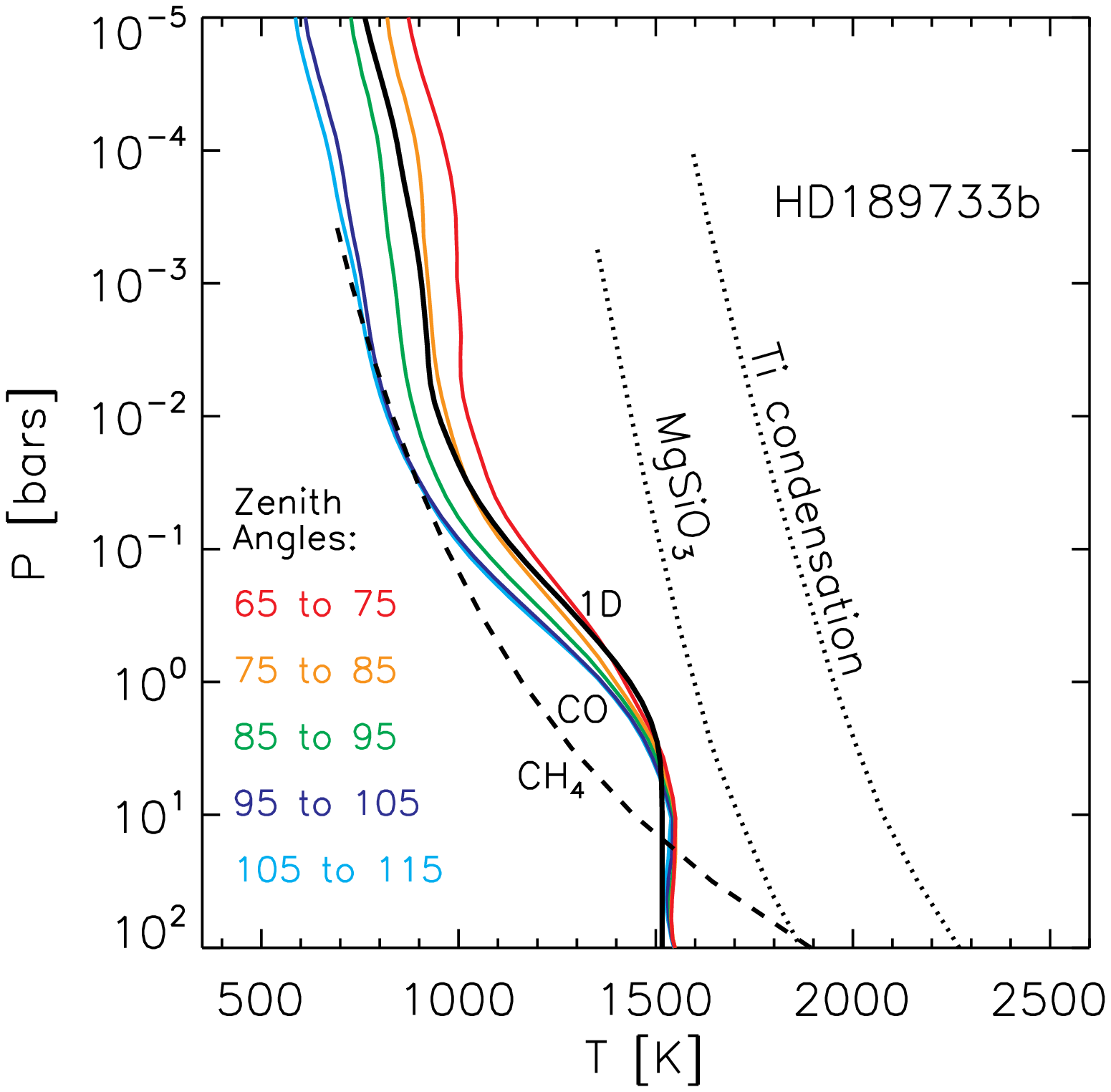}
\caption{Atmospheric \emph{P-T} profiles at the limb of \he, from the 3D model.  Smaller zenith angles are towards the day side, and larger zenith angles towards the night side.  From day-side to night-side, these are red, orange, green, blue, and cyan.  The 1D planet-wide average \emph{P-T} profile \cp{Fortney06} is shown in black.  The dashed black curve shows where CO and CH$_4$ have equal abundances.  CO is increasingly favored at higher temperatures.  The condensation curve where TiO gas is lost to solid condensates is shown as a dotted black curve, as is the condensation curve for enstatite, MgSiO$_3$.
\label{pt189}}
\end{figure}

\begin{figure}
\plotone{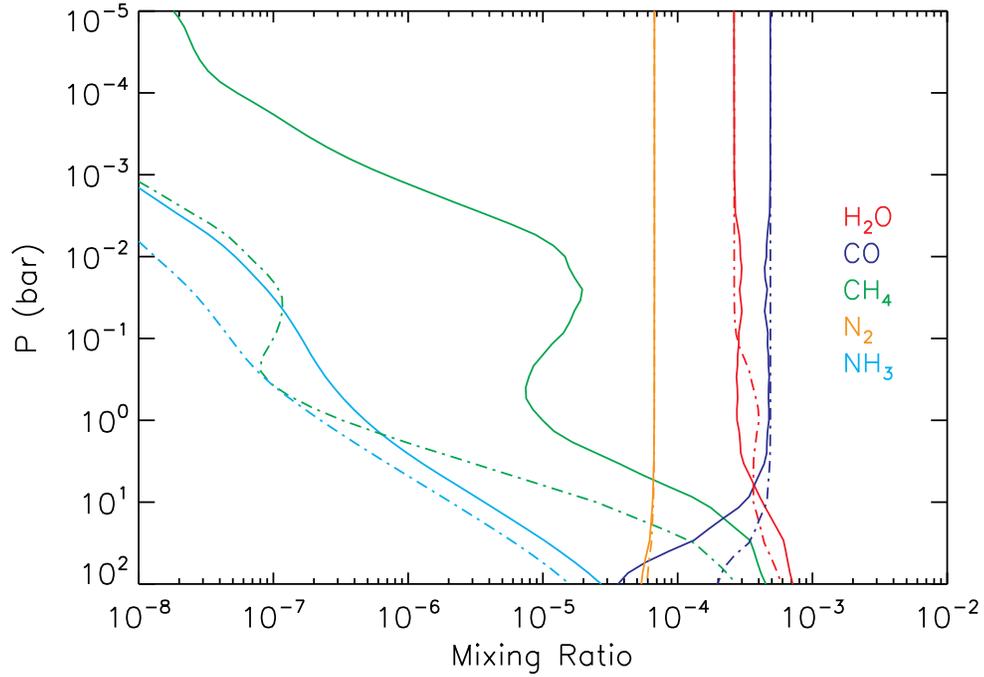}
\caption{Equilibrium chemistry mixing ratios for several molecules for the 1D profiles of \he\ (solid lines) and \hd\ (dash-dot lines).  The metallicity is solar.
\label{chem}}
\end{figure}

In \mbox{Figure \ref{pt189}} we compare an update of the \ct{Fortney06} 1D profile (in black) to average profiles (averaged over all azimuths) from the 3D model, as a function of zenith angle, near the terminator.\footnote{We note that these azimuthally averaged profiles are not used in the actual transmission spectra calculations.  We perform the transmission spectrum calculations on a column-by-column basis, as a function of position, in the atmosphere.  The azimuthally averaged profiles merely show average conditions at the planet's limb.}  The 1D model is a planet-wide average profile that assumes that the incident flux is first cut in half (for a day-night average) and then that the incident flux arrives from $\mu$=0.5 (meaning the incident angle is 60$^{\circ}$), which is the dayside average \cp[see][]{Fortney07b}.  As previously suggested, the 1D profile is a good approximation to the true limb profiles from the 3D model.  Although the CO/CH$_4$ ratio is always $>1$ for these mean profiles, given chemical equilibrium and the location of the profiles in \emph{P-T}-space, we should see CH$_4$ absorption features, in addition to CO, in the transmission spectrum, since CH$_4$ is abundant.  The equilibrium chemistry mixing ratios, for C-N-O molecules, for the 1D \he\ model are shown in \mbox{Figure \ref{chem}}.

\begin{figure}
\plotone{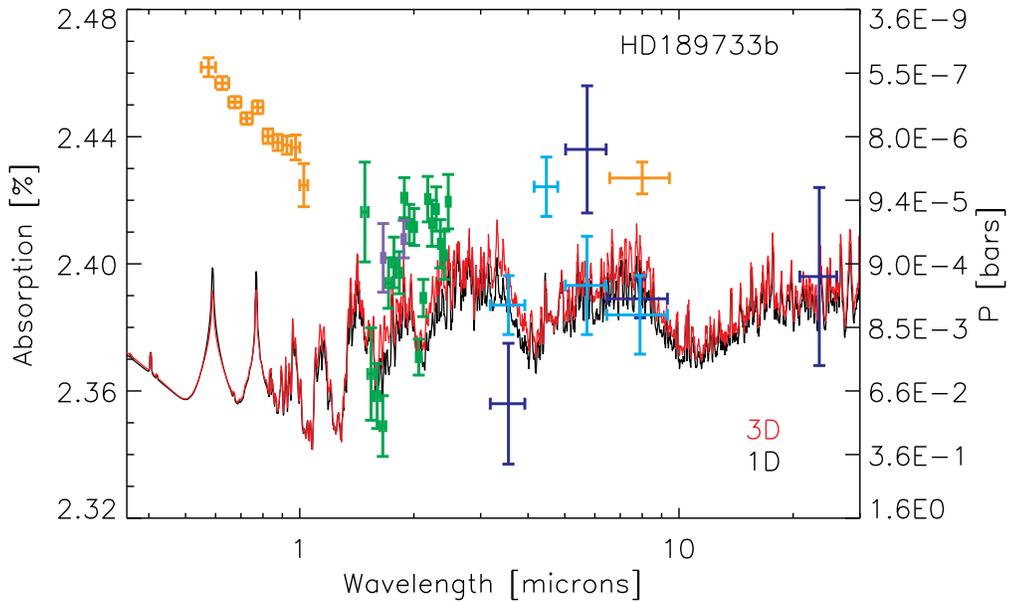}
\caption{HD 189733 in-transit absorption depth vs.~wavelength for 1D (black) and 3D (red) model atmospheres, assuming chemical equilibrium.  Computed absorption depths are (\rp / \rs)$^2$. Many data sets are also shown.  The optical (orange) is \ct{Pont08} and the near infrared (green and purple) are \ct{Swain08} and \ct{Sing09b}, respectively.  Further into the infrared, in dark blue at 3.6 and 5.8 $\mu$m are from \ct{Beaulieu08}, while the 8 $\mu$m  point is from \ct{Knutson07b}, and the 24 $\mu$m point is from \ct{Knutson09b}.  The orange 8 $\mu$m point is from \ct{Agol08}.  The four cyan IRAC points are from \ct{Desert09}.  The data of \ct{Redfield08} are of \emph{much} higher spectral resolution that the other data sets, and are not plotted.  The right y-axis shows the pressure at the terminator that corresponds to a given planet radius/absorption depth.
\label{189d}}
\end{figure}

The transmission spectrum of 1D and 3D solar-metallicity chemical equilibrium models of \he\ are shown in \mbox{Figure \ref{189d}}.  The published observations are also shown.  The right y-axis is cast as the pressure at the terminator, for the planetary radius that would cause the observed absorption depth.  The first thing to note is the similarity between the two model spectra.  Since the 1D profile is a reasonable proxy for the 3D profiles, the transmission spectra look very similar.

The second thing to notice is that the fit to some of the data are not particularly good.  The \ct{Pont08} optical data may well be due to a haze layer, perhaps due to small enstatite grains \cp{Lecavelier08a}.  This is a reasonable suggestion, given that silicate grains condense around these temperatures.  However, it is not clear why similar obscuring silicates wouldn't effect \hd, although it is possible that its atmosphere is hot enough for enstatite particles to vaporize in the upper atmosphere. (Compare Figure 5 to Figure 8.)  Abundant silicate particles at such a great altitude (pressures of $\sim$0.5 to 50 $\mu$bar to explain the observations) may indicate vigorous vertical mixing above a cloud base at $\sim$100+ bar.  Alternatively, the obscuring material could potentially be something other than silicates---perhaps a methane-derived soot \cp{Zahnle10}.

In the near-infrared, we match well the location of absorption features found by \ct{Swain08}, but not their \emph{amplitude}.  Both models are normalized to the local radius minimum at 2 $\mu$m.  What limits the amplitude of these near IR features, which are predominantly due to water vapor absorption, is a strong contribution from H$_2$ collision induced absorption (CIA) \cp[e.g.][]{Borysow02}.  The model can not reach a deeper minimum in absorption, especially at 2.1 $\mu$m, because of the CIA opacity \cp[see, e.g. Figure 15 from][]{Sharp07}.  Recently \ct{Sing09b} have analyzed transit photometry in two narrow bandpasses, over this same wavelength range, and in the bluer channel do not reproduce the depth found by \ct{Swain08}.  In the mid-infrared, it is potentially problematic to make detailed comparisons, as the \ct{Beaulieu08} and \ct{Desert09} reductions of the \emph{same} data sets do not agree.  This highlights that transmission spectrum observations are still very much on the edge of what can be done, so choices made in data reduction have direct consequences on the interpretation.  In addition, for the HD189733 system in particular, starspots can in principle make observed eclipse depths change from month to month, which may effect the comparison of models to multiple datasets taken at different times.

\begin{figure}
\plotone{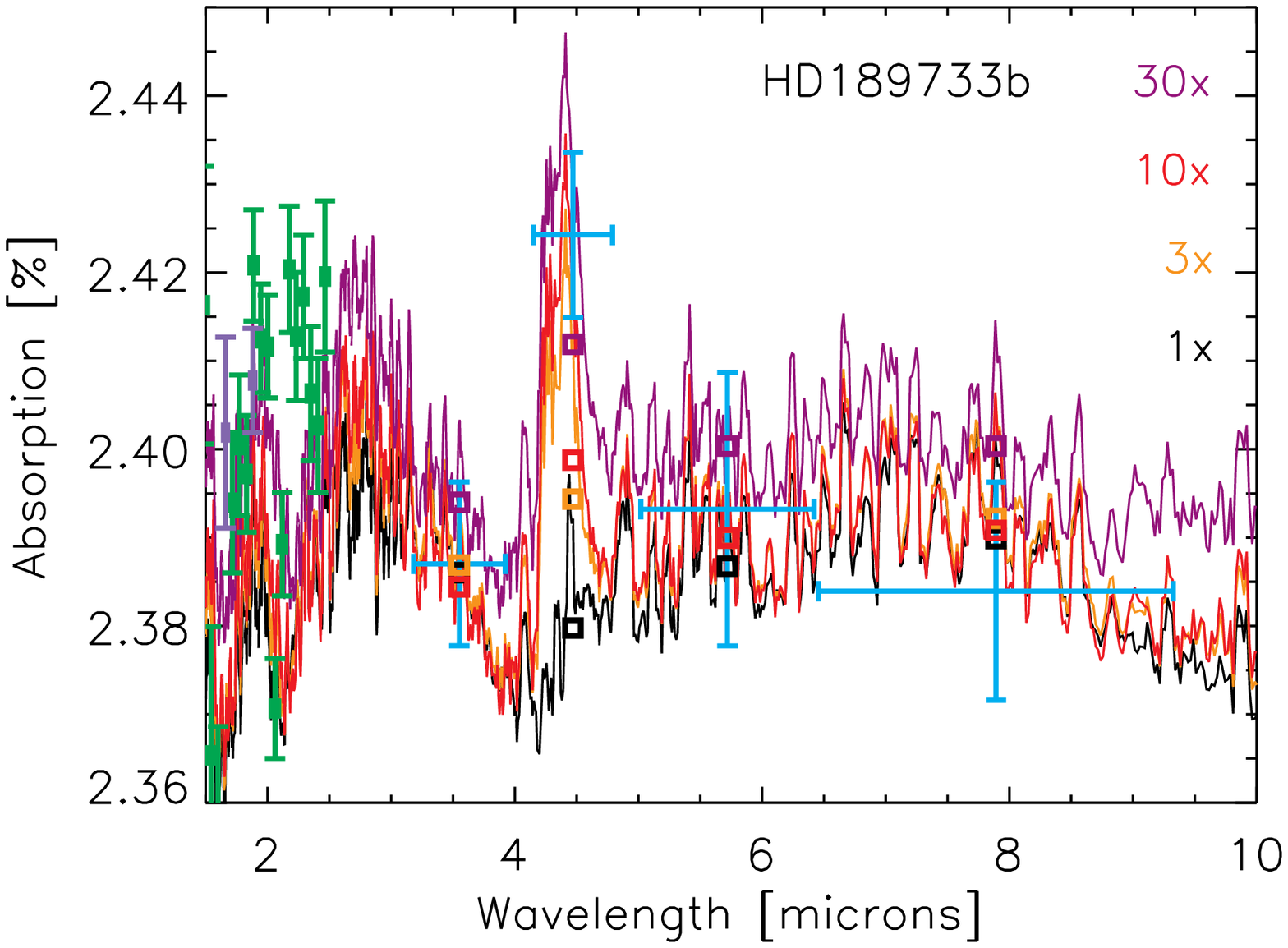}
\caption{In-transit absorption depth vs.~wavelength for 1D models of \he\ at four metallicities, from 1$\times$ to 30$\times$ solar. Computed absorption depths are (\rp / \rs)$^2$.  We compare in detail to the mid-infrared data of \ct{Desert09}, from 3-10 $\mu$m.  Band-averaged model absorption depths are shown as squares.  A higher metallicity favors CO$_2$ and CO, at the expense of CH$_4$, yielding a better fit to the 4.5$\mu$m band in particular.
\label{189met}}
\end{figure}

If we try to fit only the \ct{Desert09} data set from 3-10 $\mu$m, we basically concur with their findings that a large CO abundance could be required, given the large radius in the 4.5 $\mu$m band.  Alternately, however, CO$_2$ could be much more important than CO.  While the mixing ratio of CO increases linearly with metallicity the abundance of CO$_2$ increases quadratically \cp[e.g.][]{Zahnle09}.  Our fit to the IRAC 4.5 $\mu$m data is much better at high metallicity, as a strong CO$_2$ band at 4.4 $\mu$m increases the planetary radius, along with CO, in this band.  In \mbox{Figure \ref{189met}} we compare transmission spectra calculated at 1, 3, 10, and 30$\times$ solar, for the 1D model, with the corresponding correct MMWs, to the \ct{Desert09} data.  We also overplot the band-averaged model calculations, and find that we can fit near the 1$\sigma$ error bars with the 30$\times$ solar model.  Alternatively, it is likely possible to fit with only a solar abundance of H$_2$O, but a large CO or CO$_2$ abundance.  It is not possible for for us to achieve a fit to the \ct{Beaulieu08} mid-IR data points plotted in \mbox{Figure \ref{189d}}, with our \emph{P-T} profile, unless we artificially decrease the planet's gravity by a factor of $\sim$~5, to expand the scale height by this same factor.

These fits bring up a modeling issue that we have not been able to resolve.  In general, we are unable to match the large variation in transit radius of the models of Tinetti and colleagues.  This is true for their optical spectrum of \he\ shown in \ct{Pont08}, as well as the mid-infrared spectra for \he\ and \hd\ from \ct{Tinetti07a}, even when we use the \ct{Tinetti07a} \emph{P-T} profiles.  Of course even when using the same profiles, the chemical mixing ratios may be slightly different between our calculations and theirs, although that may not explain the dramatic differences.  Different opacity databases may be more imporant, as high opacity directly corresponds to large radii and low opacity to small radii.  We have had better success in comparisons with other authors \cp{Barman07,Sing08b,MillerRicci09a}, and we match exactly the analytic radius vs.~wavelength relation for a Rayleigh scattering atmosphere \cp{Lecavelier08b}.  Recently, \ct{Madhu09} investigated a large range of compositions for \he, and the amplitude of their absorption features may be intermediate between those of Tinnetti et al.~and our own.  Of course, data is the true constraint, and  Tinetti and colleagues have shown excellent agreement with the \ct{Swain08} and \ct{Beaulieu09} observations.  Good agreement between models with similar inputs is important for constraining model atmospheres and for deriving chemical mixing ratios in observed planets, so this issue should be addressed with some care in the near future.

\section{\hd} \label{sec209}
Significant attention has also been paid to \hd, which for many years was the only hot Jupiter amenable to atmospheric characterization.  Neutral atomic sodium was first discovered by \ct{Charb00}, and later confirmed by other authors \cp{Snellen08,Sing08a,Langland09}.  There seems to be little agreement as to the abundance of Na, and it could be time-variable.  \ct{Barman02} found that non-LTE level populations of Na could contribute to a weaker absorption feature at 589 nm, and \ct{Fortney03} found that photoionization of Na could be important.

\ct{Sing08a} used the data of \ct{Knutson07a} to produce an optical transmission spectrum from 300 to 770 nm.  The \ct{Sing08a} data show absorption via Na, potentially Rayleigh scattering (likely due to H$_2$) in the blue, and perhaps additional absorption, due to another absorber, which may be TiO and VO \cp{Desert08}.  \ct{Sing08b} suggest that the Na feature shape is best matched by an atmosphere enhanced in Na at lower pressures and depleted at lower pressures, due either to condensation of Na into solids, or photoionization.  Very recently \ct{Beaulieu09} have published 4-band IRAC photometry of the transit of \hd, which is the same mid-IR spectral coverage attained for \he\ by \ct{Desert09}. \ct{Beaulieu09} interpret the radius variation observations as being due essentially only to water vapor, similar to the work of \ct{Tinetti07} for \he.

\begin{figure}
\plotone{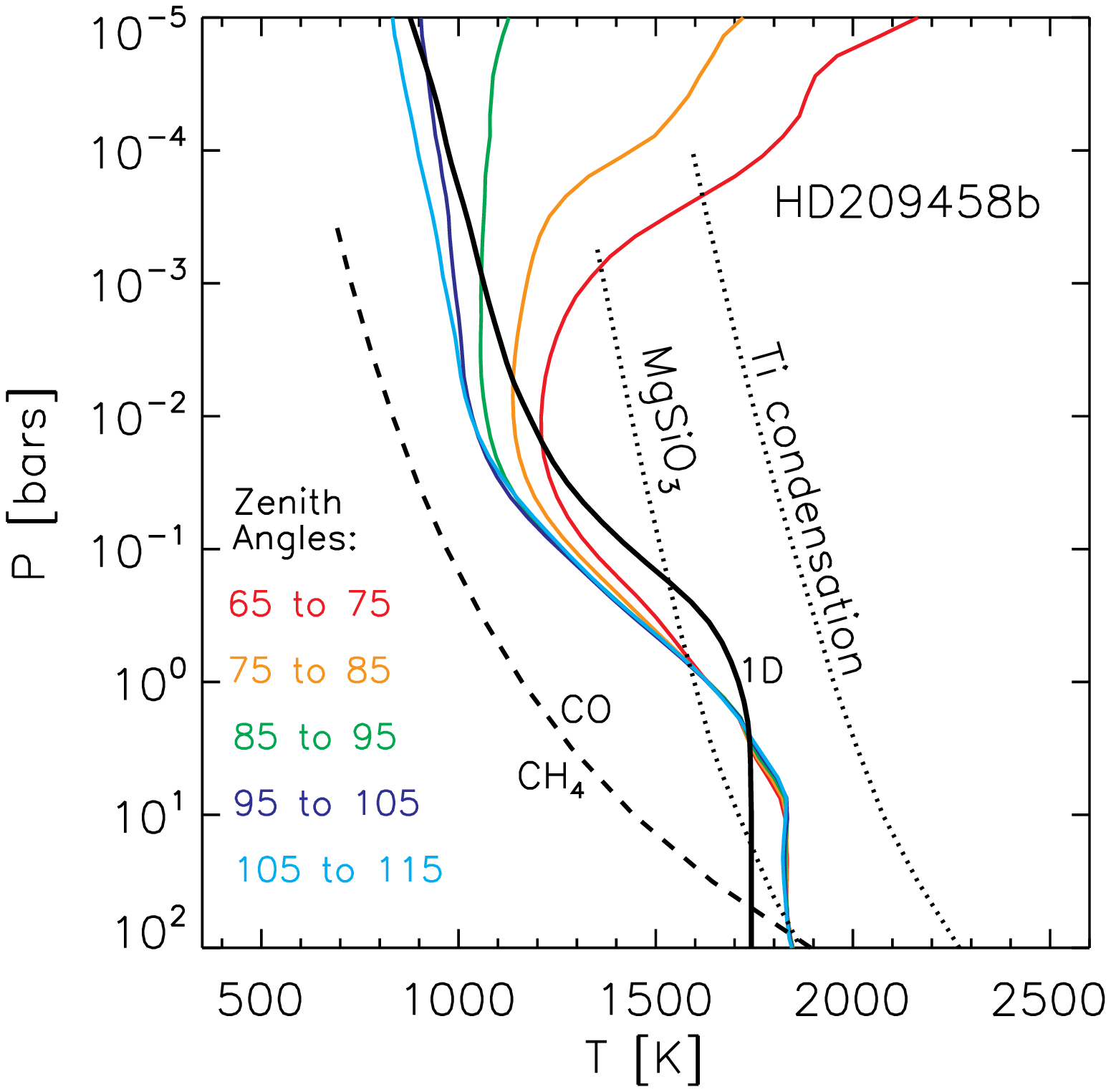}
\caption{Atmospheric \emph{P-T} profiles at the limb of \hd, from the 3D model.  Smaller zenith angles are towards the day side, and larger zenith angles towards the night side.  From day-side to night-side, these are red, orange, green, blue, and cyan.  The 1D planet-wide average \emph{P-T} profile \cp{Fortney05} is shown in black.  The dashed black curve shows where CO and CH$_4$ have equal abundances.  CO is increasingly favored at higher temperatures.  The condensation curve where TiO gas is lost to solid condensates is shown as a dotted black curve, as is the condensation curve for enstatite, MgSiO$_3$.  There is clearly a larger spread in the temperatures at the limb of \hd\ than for \he.
\label{pt209}}
\end{figure}

In Figure \ref{pt209} we plot the azimuth-averaged \emph{P-T} profiles for the 3D model of \hd, near the planet's terminator region, and compare this to our 1D planet-wide average model, updated from \ct{Fortney05}.  It is clear that \hd\ possesses a larger diversity in \emph{P-T} profiles in this region than \he.  Models on the day side of the terminator have a temperature inversion, due to heating of the stratosphere via absorption of optical flux by TiO.  The night side of the terminator region lacks TiO.  For the 1D planet-wide average profile (black), we do not find a temperature inversion, although for a day-side average model (which uses a 2$\times$ higher incident flux), we do find an inversion \cp{Fortney08a}.

\begin{figure}
\plotone{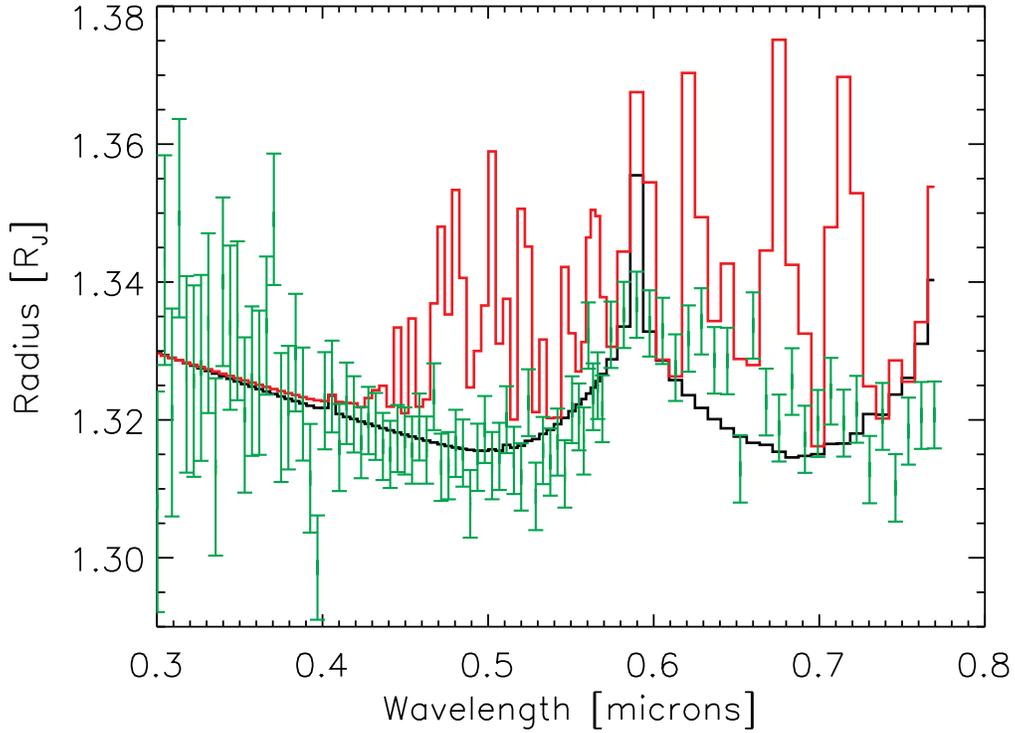}
\caption{Transit radii of \hd\ 1D (black) and 3D (red) models, compared to the data of \ct{Sing08a}, in green.  The models, which assume chemical equilibrium, have been binned onto the Sing et al.~wavelength grid.  The 1D model shows Rayleigh scattering in the blue, along with a broad Na absorption feature at 589 nm, that is stronger than observed.  The 3D model shows weak bands due to TiO and VO, that are still much stronger than observed.  The data are plotted assuming a parent star radius of 1.125 \rs, after \ct{Knutson07a}.
\label{209data}}
\end{figure}

The comparison of the 1D and 3D models with the optical data of \ct{Sing08a} in \mbox{Figure \ref{209data}} dramatically shows the difference in transmission spectra that these temperatures induce.  The 1D model, which lacks any appreciable amount of TiO or VO, generally matches the Na absorption features as well as the radius increase in the blue, here due to Rayleigh scattering off of H$_2$.  The thin core of the Na absorption feature is too strong in our model, which bolsters the finding of \ct{Sing08a} that Na is depleted in the upper atmosphere.  The 3D model also shows the broad, pressure-broadened absorption of Na and K, with numerous weak TiO/VO bands at some wavelengths.  These bands are due to a significant depletion of TiO/VO compared to solar abundances.  As shown in Figures \ref{g10} and \ref{g50}, a solar complement of TiO/VO leads to a dramatically larger radius.  The 3D model looks somewhat similar to the 1500 K (green) models in these earlier figures.

This basic picture of depleted TiO/VO is similar to the suggestion by \ct{Desert08} of weak TiO/VO bands red-ward of the Na doublet, but we do not achieve as good of a fit, since we do not leave the molecular abundances as free parameters.  A small change in temperature could well lead to a fit similar to that of \ct{Desert08}.  \ct{Spiegel09} have recently highlighted the myriad difficulties in modeling the vertical distribution of TiO in a 1D model.  It seems quite plausible that TiO, should it exist on the day side of \hd, and cause its temperature inversion, could be quite depleted at the planet's terminator, where temperatures are colder. However, the 3D model does not reproduce the observed mid IR secondary eclipse photometry \cp[see][]{Showman09} or transmission spectrum very well.

\begin{figure}
\plotone{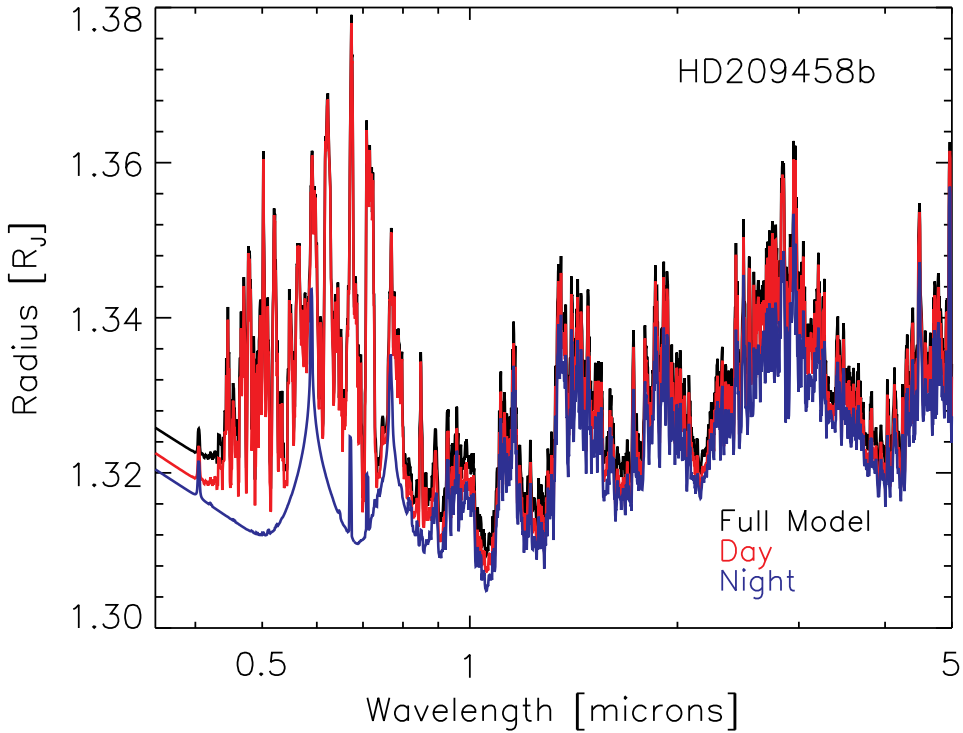}
\caption{Transmission spectrum contributions of the day (red) and night (blue) hemispheres of the planet.  The full-planet model is shown in black.  The larger scale height and higher optical opacity of the day side lead it to dominate the planet-wide transmission spectrum.
\label{daynite}}
\end{figure}

To further explore the terminator of \hd\ we also plot the transmission spectra contributions of the day and night hemispheres separately, in \mbox{Figure \ref{daynite}}.  The night hemisphere (blue) shows no TiO/VO, while the day hemisphere (red) shows the weak TiO/VO bands.  The planet-wide spectrum (black) shows that it is the day side contribution that dominates the transmission spectrum.  This is for two reasons:  First, the scale height is larger on the warmer day side.  This ``puffing up'' masks the absorption features formed on the night side.  This will likely happen for any planet.  Second, for \hd\ in the optical in particular, the opacity on the day side is far larger than that on the night side, which further masks the contribution from the night side.

\begin{figure}
\plotone{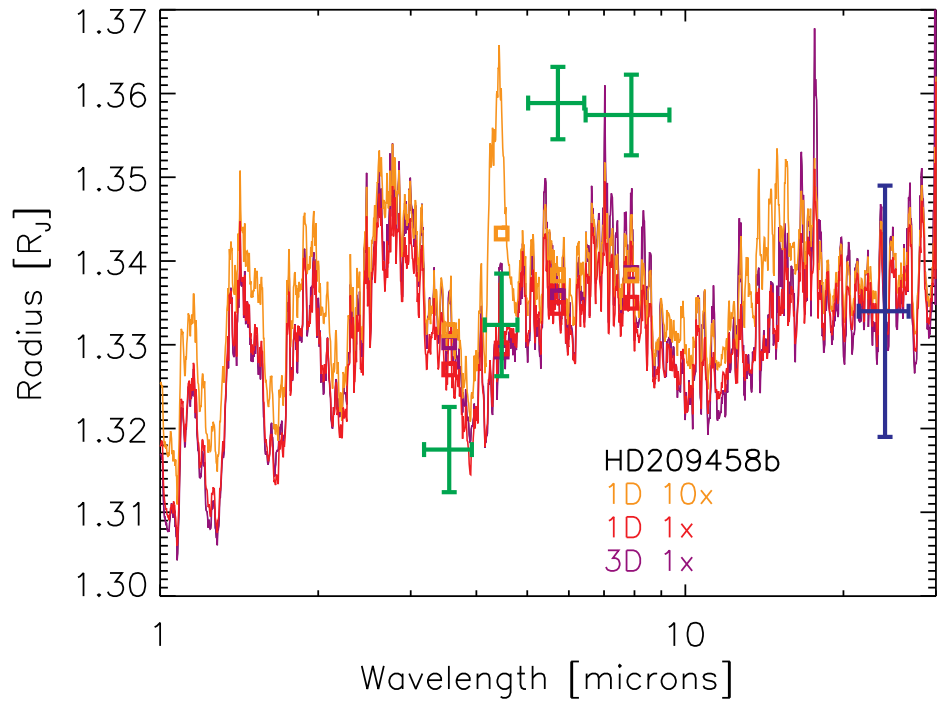}
\caption{Transit radius of \hd\ 1D and 3D models, compared to the IRAC data of \ct{Beaulieu09} and the mean 24 $\mu$m MIPS data point of \ct{Richardson06} and H.~Knutson (pers.~ comm.).  The 3D model (which is 1$\times$ solar) is shown in purple, while 1D models at 1$\times$ and 10$\times$ solar metallicity are shown in red and orange respectively.  The base radius is chosen to give a good fit to the optical observations.  Band-averaged model radii over the IRAC bandpasses are shown in squares.  The data are plotted assuming a parent star radius of 1.125 \rs, after \ct{Knutson07a}.  Water vapor and CO features dominate the mid-IR spectrum, as in \he, while the 4.4 $\mu$m feature in the 10$\times$ model is due to CO$_2$.
\label{209irac}}
\end{figure}

We next compare our models to the recent 4-band IRAC data on the transit of \hd\ \cp{Beaulieu09}.  In \mbox{Figure \ref{209irac}} we compare the 3D model at solar metallicity, along with the 1D model at 1$\times$ and 10$\times$ solar, to the mid-IR data set.  Here the base 10-bar radius is chosen to give a good match to the data in the optical for the 1D models, and this same base radius is used for the 3D model.  While the solar metallicity models generally reproduce the slope towards larger radii at longer wavelength, none of the models reproduce the large radius variations found by \ct{Beaulieu09}.  Together with the previously published poor fit of the \hd\ 3D model to the secondary eclipse data \cp{Showman09}, this may indicate that the model does not reproduce the true mid-infrared opacity of the planet.  We also note that while the 1D and 3D models have dramatically different optical transit radii, due to the temperature-sensitive TiO/VO abundances, the same is not true in the mid-infrared, where both the 1D and 3D solar-metallicity models are dominated my H$_2$O and CO opacity, and predict very similar radii.

\section{Spectral Differences of the Trailing and Leading Hemispheres}
In the current era it is extremely difficult for observers to even \emph{achieve} data on hot Jupiter atmospheres.  However, in the future this will not always be the case.  We are quite optimistic about the era of the \emph{James Webb Space Telescope}, which is scheduled to launch in 2014.  It will then be possible to derive more detailed information about these planets' atmospheres.  In particular, transmission spectra of a planet's leading hemisphere (probed during the first half of transit ingress) and trailing hemisphere (during the last half of transit egress) may allow for detection of spectral differences on different locations of the planet.  This could be complementary to eclipse mapping during the secondary eclipse ingress and egress phases \cp{Williams06,Rauscher07b}.

It is a general outcome of our published 3D hot Jupiter simulations that wide low-latitude west-to-east winds persist on arbitrarily long time scales.  The first observational consequence is that the hottest point of the planet is blown downstream from the substellar point \cp{Showman02,CS05,Showman09}, which was confirmed for \he\ by \ct{Knutson07b}.  The effect is more dramatic in simulations of \he\ than for \hd, owing mostly to a longer radiative time constant in \he\ \cp{Showman08,Fortney08a}.  At the time of primary transit, this leads to the leading hemisphere being colder than the trailing hemisphere, due to the downwind displacement of both the hottest and coldest points of the planet \cp{Showman09}.

\begin{figure}
\epsscale{.75}
\plotone{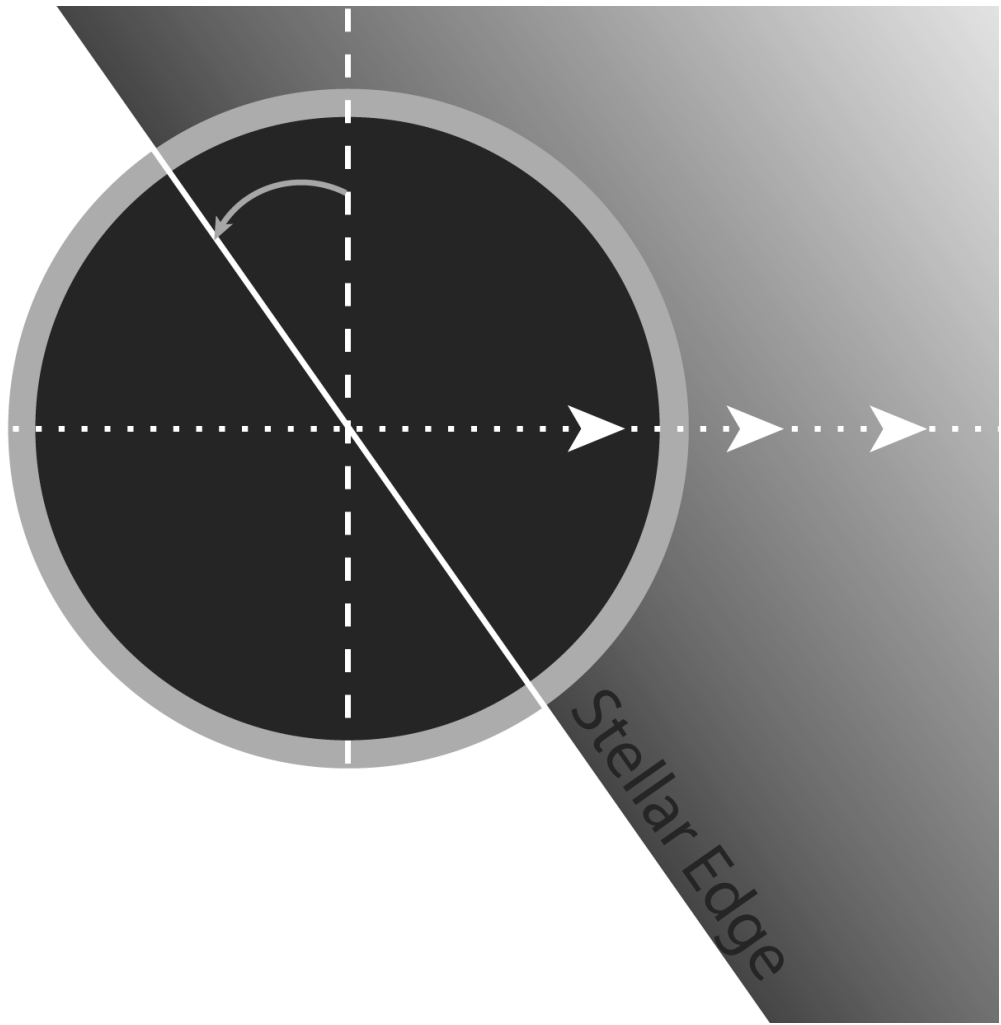}
\caption{Schematic view of ingress at a non-zero impact parameter.  Half-way through ingress, one samples mostly the leading hemisphere, but also some of the trailing hemisphere, near the pole.  The planet's rotation axis is the long-dashed line, while the planet's orbital path is the dotted line, with arrows.
\label{ingress}}
\end{figure}

This dichotomy could potentially be probed with transmission spectra obtained during the transit ingress and egress phases.  A transmission spectrum taken half-way into the time of ingress would sample only the leading hemisphere, and at egress, the trailing.  The simple leading--trailing image is complicated by the transit impact parameter, $b$, however, as shown in \mbox{Figure \ref{ingress}}.  As $b$ increases from zero (central transit) to one (grazing), the edge of the parent star becomes gradually more inclined to the planet's rotation axis.  For \he\ this leads to $38.8^{\circ}$ of trailing hemisphere (near the pole) on the leading edge, and for \hd, $30.9^{\circ}$.  In general this would serve to mute differences between the hemispheres.

\begin{figure}
\plotone{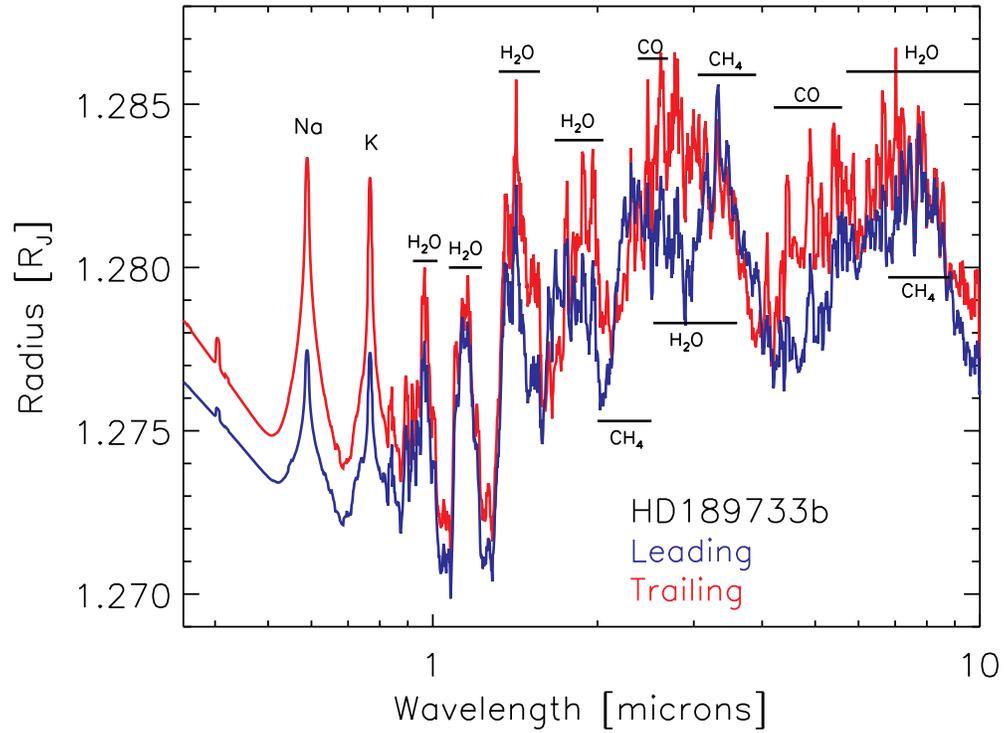}
\caption{Planet radius vs.~wavelength for the leading and trailing hemispheres of a 3D simulation of planet \he, assuming chemical equilibrium.  The leading hemisphere (blue) is cooler, while the trailing hemisphere (red) is warmer.  Cooler regions have a smaller CO/CH$_4$ ratio, which leads to changes in spectral features on the two hemispheres.  The leading hemispheres shows much stronger CH$_4$ absorption.
\label{189lead}}
\end{figure}

The differences in the spectra of the hemispheres can be seen most clearly in \he\, which is a relatively cool hot Jupiter.  Its planet-wide mean \emph{P-T} profile is close to the boundary where CO and CH$_4$ have equal abundances \cp{Fortney06}, given chemical equilibrium.  We find that if this equilibrium holds, the night side should be CH$_4$ dominated, while the day side is CO dominated \cp{Showman09}.  We have computed separate transmission spectra of the leading and trailing hemispheres (including the impact parameter effect) which are shown in \mbox{Figure \ref{189lead}}.  The differences in temperature lead to dramatic differences in spectra.  The leading (cooler) hemisphere has a smaller scale height, which leads to a smaller transit radius at most wavelengths.  However, in CH$_4$ bands, absorption is much stronger, which can lead to radii larger than that of the trailing hemisphere.  The dramatic differences in spectra may be within reach of \emph{JWST}, if multiple transits are observed (T.~Greene, personal communication).

\begin{figure}
\plotone{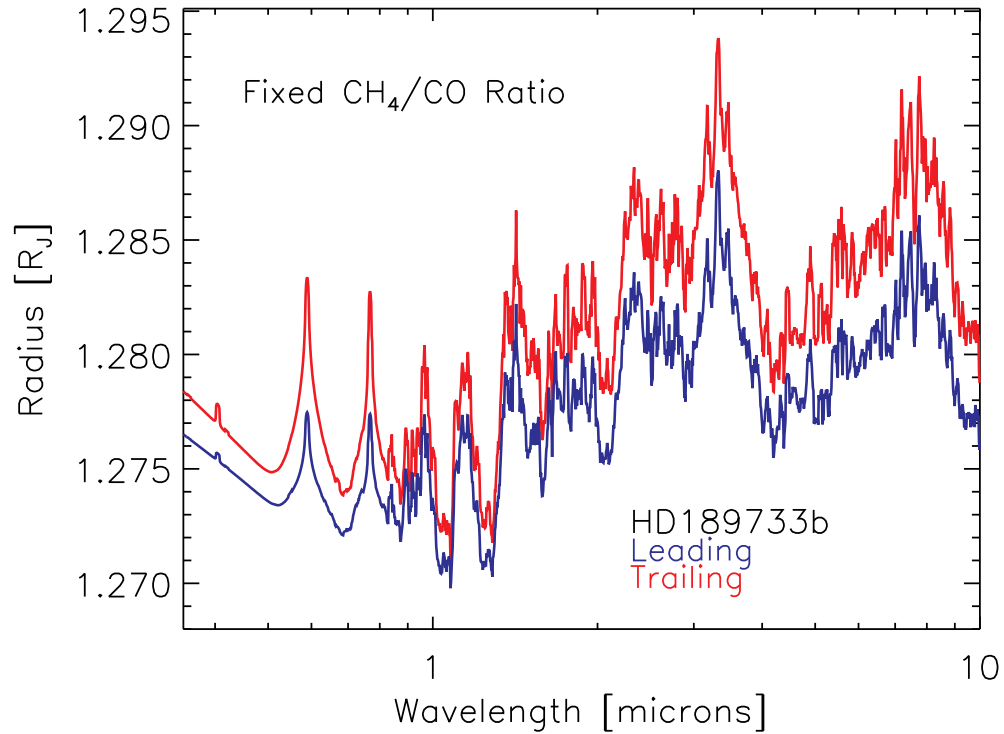}
\caption{Planet radius vs.~wavelength for the leading and trailing hemispheres of a 3D simulation of planet \he, with a fixed CO/CH$_4$ mixing ratio of 4.5/1, as could be expected from homogenization due to vertical mixing.  The leading hemisphere (blue) is cooler, while the trailing hemisphere (red) is warmer.  Differences between the hemispheres are due only to scale height differences, and the small temperature-dependent changes in opacity.
\label{189leadfix}}
\end{figure}

These CO/CH$_4$ absorption features could also be a direct probe of non-equilibrium chemistry in the atmosphere of \he, since vigorous vertical mixing would tend to homogenize the CO/CH$_4$ ratio everywhere in the planet's atmosphere \cp[see][and references therein]{CS06}.  In order to simulate how a fixed CO/CH$_4$ ratio, independent of height and location, would effect these spectra, we have calculated transmission spectra with mixing ratios given by the ``cold model'' of \ct{CS06}, which is a good approximation for the temperatures of \he.  That model yields a CO/CH$_4$ ratio of 4.5/1, meaning that CO dominates, but both molecules are abundant and should show strong absorption features.  The results are shown in \mbox{Figure \ref{189leadfix}}.  The spectra of the hemispheres are essentially identical, except for the larger scale height on the trailing hemisphere.  Absorption features of both CO and CH$_4$ can be seen.

\begin{figure}
\plotone{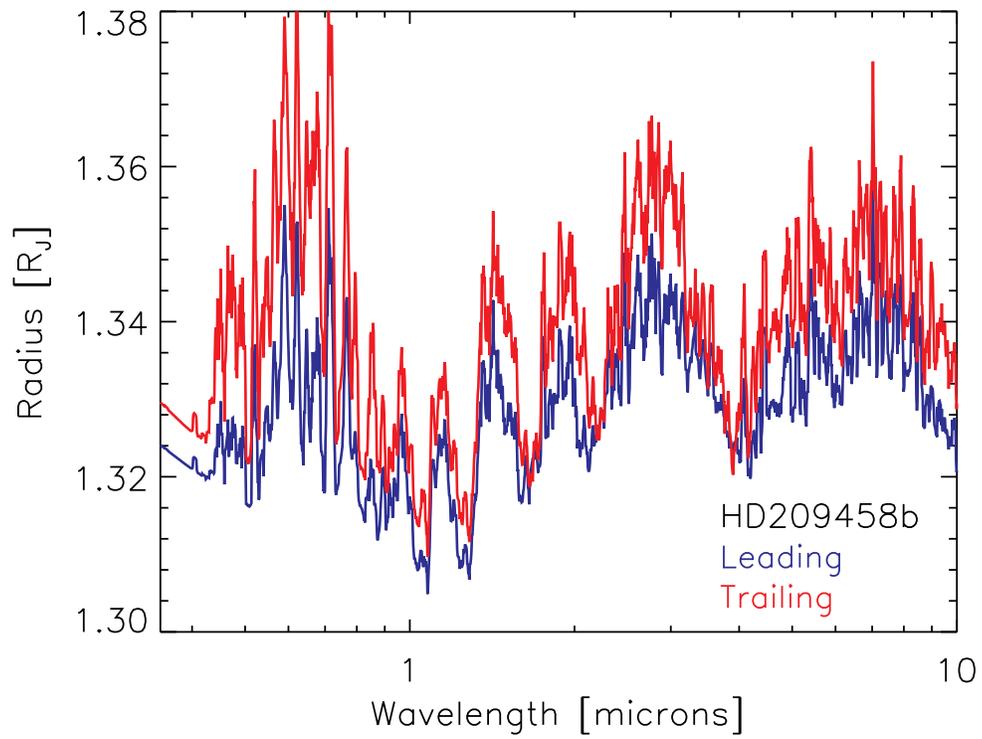}
\caption{Planet radius vs.~wavelength for of the leading and trailing hemispheres of a 3D simulation of planet \hd, assuming chemical equilibrium.  The leading hemisphere (blue) is cooler, while the trailing hemisphere (red) is warmer.  Most the differences are due to the change in the scale height, while the TiO/VO band are stronger on the warmer, trailing, hemisphere.
\label{209lead}}
\end{figure}

For \hd, which is significantly hotter than \he, we find smaller differences between the leading and trailing hemispheres.  The warmer temperatures lead both hemispheres to be CO-dominated, with CH$_4$ absorption features absent.  In addition, the shorter radiative time constant in the pM class atmosphere leads to smaller shift in the hot spot downstream \cp[see][]{Showman09}. \mbox{Figure \ref{209lead}} shows difference in the hemispheres are mostly due to small scale height differences, with some subtle effects on TiO absorption bands, due to its more depleted abundance on the cooler (leading) hemisphere.  Previously \ct{Iro05} showed a similar effect for a 3D model of \hd\ that included radiative transfer and parameterized winds.  This model lacked TiO, and they found a strong Na absorption feature at 589 nm on the hotter trailing hemisphere, but a weaker Na feature on the cooler leading hemisphere, due to the condensation of Na into clouds.  They did not investigate transmission spectra at other wavelengths.

\section{Conclusions}
We have investigated several aspects of the transmission spectra of transiting planets.  As previously known, high temperature and low surface gravity planets yield the best prospects for obtaining transmission spectra.  We have investigated the spectra for a model planet at an atmospheric metallicity of up to 30$\times$ solar, which is likely representative of Neptune-class exoplanets, and demonstrated that enhanced MMW weight in these atmospheres does not significantly diminish the magnitude of absorption features in these atmospheres.  The small cross-sectional area of the atmospheres of these smaller planets is a greater challenge.

We have calculated the transmission spectrum of 3D models of the atmospheres of \he\ and \hd, and compared these results to those of the 1D calculations.  Differences between the 3D and 1D models can be dramatic, or negligible, depending on the wavelength of observation, and the temperatures probed at the planet's terminator region.  For the planet-wide 3D transmission spectrum of \he, we find it is well reproduced by the 1D model.  Our model constrains the postulated upper haze layer \cp{Pont08} to be optically thick at optical wavelengths at microbar pressures.  Following \ct{Fortney05c}, we find that the slant optical depth at 1 $\mu$bar of an absorbing layer that shares the gaseous scale height is 60 times larger than the normal optical depth.  It remains possible that such an absorber could remain undetected in scattered light.

We are able to reproduce the IRAC transmission spectrum points of \ct{Desert09} with a large abundance of CO, and especially CO$_2$, in these models due to a large supersolar metallicity.  This is at odds with the determination of subsolar metallicity for this planet by comparison to  near-infrared data \cp{Swain08,Madhu09}.  The published 3D model of \he\ from \ct{Showman09} has been able to reproduce most of the secondary eclipse photometry and spectroscopy, the broad features of the 8 and 24 $\mu$m half-orbit light curves \cp{Knutson07b,Knutson09b}, and some features of the transmission spectrum (although an unmodeled haze apparently dominates in the optical and possibly the near IR).  This gives us confidence that these models, which are still in their infancy, are on the right path towards being a realistic simulation of dynamic hot Jupiter atmospheres.

For planet \hd, the differences between 3D and 1D models of the transmission spectrum are significant at optical wavelengths, due to the wide range of temperature probed in the upper atmosphere of \hd\ at the terminator region.  In the 3D model we find TiO in abundance on the day side \cp{Showman09} and a reduced abundance on the planet's limb.  This TiO masks absorption due to Na and K \cp[which could well be photoionized---e.g.][]{Fortney03}, and we find a poor match to the transmission spectrum of \ct{Sing08a}.  The higher temperatures and higher opacity of the day-side of the limb mask absorption features from the night side of the limb.  The 1D model, a planet-wide average, shows no TiO absorption features, and is a better match to the optical data.  The poor match to the transmissions spectrum, and to secondary eclipse photometry \cp{Showman09} indicates that the 3D model presented here and in \ct{Showman09}, which assumes chemical equilibrium and produces a day-side temperature inversion using TiO and VO, is not a good representation of the atmosphere of \hd.  It is possible that the absorbers causing the temperature inversion in this atmosphere do not have the same optical properties and/or vertical profile of the TiO gas employed in our model.  Future 3D models will explore the effects of disequilibrium compounds, such as sulfur species \cp{Zahnle09}, in producing thermal inversions.

Transmission spectra will continue to be an important tool in characterizing the atmospheres of transiting planets, with \emph{HST} in the UV to near IR, as well as with \emph{JWST} in the red optical to mid IR.  Probing the atmospheres with this method will continue to be complementary to secondary eclipse spectra, since different regions of the atmosphere (in latitude, longitude, and height) are probed with each method.  In addition, transmission spectra are less sensitive to the temperature gradient, while this is a major complication for emission spectra \cp[e.g.][]{Fortney06b}.

One could imagine more challenging levels of atmospheric characterization, such as the detection of wind speeds via the Doppler shift of absorption lines in the planetary atmosphere \cp{Brown01}.  This could potentially be done during the ingress and egress phases.  However, as discussed by \ct{Spiegel07}, the characterization of transiting planet atmospheres at this level requires the consideration of a number of additional effects, such as the planet's rotation rate and orbital motion, amongst others.  We will consider this in the future.

\acknowledgements
We thank the referee for comments that helped to clarify our presentation.  We thank Tom Greene, David Sing, Jean-Michel D{\'e}sert, Eliza Miller-Ricci, Travis Barman, Gautam Vasisht, Mark Swain, Jean-Philippe Beaulieu, and Giovanna Tinetti for insightful discussions.
J.~J.~F.~acknowledges the support of a \emph{Spitzer} Theory Program grant.  A.~P.~S. acknowledges support from the Origins and Spitzer programs.  M.~S.~M.~acknowledges the support of the NASA Planetary Atmospheres Program. 


\begin{thebibliography}{92}
\expandafter\ifx\csname natexlab\endcsname\relax\def\natexlab#1{#1}\fi

\bibitem[{{Adcroft} {et~al.}(2004){Adcroft}, {Campin}, {Hill}, \&
  {Marshall}}]{Adcroft04}
{Adcroft}, A., {Campin}, J.-M., {Hill}, C., \& {Marshall}, J. 2004, Monthly
  Weather Review, 132, 2845

\bibitem[{{Agol} {et~al.}(2009){Agol}, {Cowan}, {Bushong}, {Knutson},
  {Charbonneau}, {Deming}, \& {Steffen}}]{Agol08}
{Agol}, E., {Cowan}, N.~B., {Bushong}, J., {Knutson}, H., {Charbonneau}, D.,
  {Deming}, D., \& {Steffen}, J.~H. 2009, in IAU Symposium, Vol. 253, 209--215

\bibitem[{{Barman}(2007)}]{Barman07}
{Barman}, T. 2007, \apjl, 661, L191

\bibitem[{{Barman} {et~al.}(2001){Barman}, {Hauschildt}, \&
  {Allard}}]{Barman01}
{Barman}, T.~S., {Hauschildt}, P.~H., \& {Allard}, F. 2001, \apj, 556, 885

\bibitem[{{Barman} {et~al.}(2005){Barman}, {Hauschildt}, \&
  {Allard}}]{Barman05}
---. 2005, \apj, 632, 1132

\bibitem[{{Barman} {et~al.}(2002){Barman}, {Hauschildt}, {Schweitzer},
  {Stancil}, {Baron}, \& {Allard}}]{Barman02}
{Barman}, T.~S., {Hauschildt}, P.~H., {Schweitzer}, A., {Stancil}, P.~C.,
  {Baron}, E., \& {Allard}, F. 2002, \apjl, 569, L51

\bibitem[{{Beaulieu} {et~al.}(2008){Beaulieu}, {Carey}, {Ribas}, \&
  {Tinetti}}]{Beaulieu08}
{Beaulieu}, J.~P., {Carey}, S., {Ribas}, I., \& {Tinetti}, G. 2008, \apj, 677,
  1343

\bibitem[{{Beaulieu} {et~al.}(2009){Beaulieu}, {Kipping}, {Batista}, {Tinetti},
  {Ribas}, {Carey}, {Noriega-Crespo}, {Griffith}, {Campanella}, {Dong},
  {Tennyson}, {Barber}, {Deroo}, {Fossey}, {Liang}, {Swain}, {Yung}, \&
  {Allard}}]{Beaulieu09}
{Beaulieu}, J.~P., {Kipping}, D.~M., {Batista}, V., {Tinetti}, G., {Ribas}, I.,
  {Carey}, S., {Noriega-Crespo}, J.~A., {Griffith}, C.~A., {Campanella}, G.,
  {Dong}, S., {Tennyson}, J., {Barber}, R.~J., {Deroo}, P., {Fossey}, S.~J.,
  {Liang}, D., {Swain}, M.~R., {Yung}, Y., \& {Allard}, N. 2009, MNRAS in
  press, ArXiv:0909.0185

\bibitem[{{B{\'e}zard} {et~al.}(2002){B{\'e}zard}, {Lellouch}, {Strobel},
  {Maillard}, \& {Drossart}}]{Bezard02}
{B{\'e}zard}, B., {Lellouch}, E., {Strobel}, D., {Maillard}, J.-P., \&
  {Drossart}, P. 2002, Icarus, 159, 95

\bibitem[{{Borysow}(2002)}]{Borysow02}
{Borysow}, A. 2002, \aap, 390, 779

\bibitem[{{Bouchy} {et~al.}(2005){Bouchy}, {Udry}, {Mayor}, {Moutou}, {Pont},
  {Iribarne}, {da Silva}, {Ilovaisky}, {Queloz}, {Santos}, {S{\'e}gransan}, \&
  {Zucker}}]{Bouchy05}
{Bouchy}, F., {Udry}, S., {Mayor}, M., {Moutou}, C., {Pont}, F., {Iribarne},
  N., {da Silva}, R., {Ilovaisky}, S., {Queloz}, D., {Santos}, N.~C.,
  {S{\'e}gransan}, D., \& {Zucker}, S. 2005, \aap, 444, L15

\bibitem[{{Brown}(2001)}]{Brown01}
{Brown}, T.~M. 2001, \apj, 553, 1006

\bibitem[{{Burrows} {et~al.}(2008){Burrows}, {Budaj}, \& {Hubeny}}]{Burrows08}
{Burrows}, A., {Budaj}, J., \& {Hubeny}, I. 2008, \apj, 678, 1436

\bibitem[{{Burrows} {et~al.}(2007){Burrows}, {Hubeny}, {Budaj}, {Knutson}, \&
  {Charbonneau}}]{Burrows07c}
{Burrows}, A., {Hubeny}, I., {Budaj}, J., {Knutson}, H.~A., \& {Charbonneau},
  D. 2007, \apjl, 668, L171

\bibitem[{{Burrows} {et~al.}(2005){Burrows}, {Hubeny}, \&
  {Sudarsky}}]{Burrows05b}
{Burrows}, A., {Hubeny}, I., \& {Sudarsky}, D. 2005, \apjl, 625, L135

\bibitem[{{Charbonneau} {et~al.}(2005){Charbonneau}, {Allen}, {Megeath},
  {Torres}, {Alonso}, {Brown}, {Gilliland}, {Latham}, {Mandushev}, {O'Donovan},
  \& {Sozzetti}}]{Charb05}
{Charbonneau}, D., {Allen}, L.~E., {Megeath}, S.~T., {Torres}, G., {Alonso},
  R., {Brown}, T.~M., {Gilliland}, R.~L., {Latham}, D.~W., {Mandushev}, G.,
  {O'Donovan}, F.~T., \& {Sozzetti}, A. 2005, ApJ, 626, 523

\bibitem[{{Charbonneau} {et~al.}(2000){Charbonneau}, {Brown}, {Latham}, \&
  {Mayor}}]{Charb00}
{Charbonneau}, D., {Brown}, T.~M., {Latham}, D.~W., \& {Mayor}, M. 2000, \apjl,
  529, L45

\bibitem[{{Charbonneau} {et~al.}(2002){Charbonneau}, {Brown}, {Noyes}, \&
  {Gilliland}}]{Charb02}
{Charbonneau}, D., {Brown}, T.~M., {Noyes}, R.~W., \& {Gilliland}, R.~L. 2002,
  \apj, 568, 377

\bibitem[{{Charbonneau} {et~al.}(2008){Charbonneau}, {Knutson}, {Barman},
  {Allen}, {Mayor}, {Megeath}, {Queloz}, \& {Udry}}]{Charb08}
{Charbonneau}, D., {Knutson}, H.~A., {Barman}, T., {Allen}, L.~E., {Mayor}, M.,
  {Megeath}, S.~T., {Queloz}, D., \& {Udry}, S. 2008, \apj, 686, 1341

\bibitem[{{Cooper} \& {Showman}(2005)}]{CS05}
{Cooper}, C.~S. \& {Showman}, A.~P. 2005, \apjl, 629, L45

\bibitem[{{Cooper} \& {Showman}(2006)}]{CS06}
---. 2006, \apj, 649, 1048

\bibitem[{{D{\'e}sert} {et~al.}(2009){D{\'e}sert}, {Lecavelier des Etangs},
  {H{\'e}brard}, {Sing}, {Ehrenreich}, {Ferlet}, \& {Vidal-Madjar}}]{Desert09}
{D{\'e}sert}, J.-M., {Lecavelier des Etangs}, A., {H{\'e}brard}, G., {Sing},
  D.~K., {Ehrenreich}, D., {Ferlet}, R., \& {Vidal-Madjar}, A. 2009, \apj, 699,
  478

\bibitem[{{D{\'e}sert} {et~al.}(2008){D{\'e}sert}, {Vidal-Madjar}, {Lecavelier
  Des Etangs}, {Sing}, {Ehrenreich}, {H{\'e}brard}, \& {Ferlet}}]{Desert08}
{D{\'e}sert}, J.-M., {Vidal-Madjar}, A., {Lecavelier Des Etangs}, A., {Sing},
  D., {Ehrenreich}, D., {H{\'e}brard}, G., \& {Ferlet}, R. 2008, \aap, 492, 585

\bibitem[{{Dobbs-Dixon} \& {Lin}(2008)}]{Dobbs08}
{Dobbs-Dixon}, I. \& {Lin}, D.~N.~C. 2008, \apj, 673, 513

\bibitem[{{Ehrenreich} {et~al.}(2007){Ehrenreich}, {H{\'e}brard}, {Lecavelier
  des Etangs}, {Sing}, {D{\'e}sert}, {Bouchy}, {Ferlet}, \&
  {Vidal-Madjar}}]{Ehrenreich07}
{Ehrenreich}, D., {H{\'e}brard}, G., {Lecavelier des Etangs}, A., {Sing},
  D.~K., {D{\'e}sert}, J.-M., {Bouchy}, F., {Ferlet}, R., \& {Vidal-Madjar}, A.
  2007, \apjl, 668, L179

\bibitem[{{Fortney}(2005)}]{Fortney05c}
{Fortney}, J.~J. 2005, \mnras, 364, 649

\bibitem[{{Fortney} {et~al.}(2006{\natexlab{a}}){Fortney}, {Cooper}, {Showman},
  {Marley}, \& {Freedman}}]{Fortney06b}
{Fortney}, J.~J., {Cooper}, C.~S., {Showman}, A.~P., {Marley}, M.~S., \&
  {Freedman}, R.~S. 2006{\natexlab{a}}, \apj, 652, 746

\bibitem[{{Fortney} {et~al.}(2008){Fortney}, {Lodders}, {Marley}, \&
  {Freedman}}]{Fortney08a}
{Fortney}, J.~J., {Lodders}, K., {Marley}, M.~S., \& {Freedman}, R.~S. 2008,
  \apj, 678, 1419

\bibitem[{{Fortney} \& {Marley}(2007)}]{Fortney07b}
{Fortney}, J.~J. \& {Marley}, M.~S. 2007, \apjl, 666, L45

\bibitem[{{Fortney} {et~al.}(2005){Fortney}, {Marley}, {Lodders}, {Saumon}, \&
  {Freedman}}]{Fortney05}
{Fortney}, J.~J., {Marley}, M.~S., {Lodders}, K., {Saumon}, D., \& {Freedman},
  R. 2005, \apjl, 627, L69

\bibitem[{{Fortney} {et~al.}(2006{\natexlab{b}}){Fortney}, {Saumon}, {Marley},
  {Lodders}, \& {Freedman}}]{Fortney06}
{Fortney}, J.~J., {Saumon}, D., {Marley}, M.~S., {Lodders}, K., \& {Freedman},
  R.~S. 2006{\natexlab{b}}, \apj, 642, 495

\bibitem[{{Fortney} {et~al.}(2003){Fortney}, {Sudarsky}, {Hubeny}, {Cooper},
  {Hubbard}, {Burrows}, \& {Lunine}}]{Fortney03}
{Fortney}, J.~J., {Sudarsky}, D., {Hubeny}, I., {Cooper}, C.~S., {Hubbard},
  W.~B., {Burrows}, A., \& {Lunine}, J.~I. 2003, \apj, 589, 615

\bibitem[{{Freedman} {et~al.}(2008){Freedman}, {Marley}, \&
  {Lodders}}]{Freedman08}
{Freedman}, R.~S., {Marley}, M.~S., \& {Lodders}, K. 2008, \apjs, 174, 504

\bibitem[{{Geballe} {et~al.}(2009){Geballe}, {Saumon}, {Golimowski}, {Leggett},
  {Marley}, \& {Noll}}]{Geballe09}
{Geballe}, T.~R., {Saumon}, D., {Golimowski}, D.~A., {Leggett}, S.~K.,
  {Marley}, M.~S., \& {Noll}, K.~S. 2009, \apj, 695, 844

\bibitem[{{Goody} {et~al.}(1989){Goody}, {West}, {Chen}, \& {Crisp}}]{Goody89}
{Goody}, R., {West}, R., {Chen}, L., \& {Crisp}, D. 1989, Journal of
  Quantitative Spectroscopy and Radiative Transfer, 42, 539

\bibitem[{{Henry} {et~al.}(2000){Henry}, {Marcy}, {Butler}, \&
  {Vogt}}]{Henry00}
{Henry}, G.~W., {Marcy}, G.~W., {Butler}, R.~P., \& {Vogt}, S.~S. 2000, \apjl,
  529, L41

\bibitem[{{Hubbard} {et~al.}(2001){Hubbard}, {Fortney}, {Lunine}, {Burrows},
  {Sudarsky}, \& {Pinto}}]{Hubbard01}
{Hubbard}, W.~B., {Fortney}, J.~J., {Lunine}, J.~I., {Burrows}, A., {Sudarsky},
  D., \& {Pinto}, P. 2001, \apj, 560, 413

\bibitem[{{Hubeny} \& {Burrows}(2007)}]{Hubeny07}
{Hubeny}, I. \& {Burrows}, A. 2007, \apj, 669, 1248

\bibitem[{{Iro} {et~al.}(2005){Iro}, {Bezard}, \& {Guillot}}]{Iro05}
{Iro}, N., {Bezard}, B., \& {Guillot}, T. 2005, \aap, 436, 719

\bibitem[{{Knutson} {et~al.}(2008){Knutson}, {Charbonneau}, {Allen}, {Burrows},
  \& {Megeath}}]{Knutson08}
{Knutson}, H.~A., {Charbonneau}, D., {Allen}, L.~E., {Burrows}, A., \&
  {Megeath}, S.~T. 2008, \apj, 673, 526

\bibitem[{{Knutson} {et~al.}(2007{\natexlab{a}}){Knutson}, {Charbonneau},
  {Allen}, {Fortney}, {Agol}, {Cowan}, {Showman}, {Cooper}, \&
  {Megeath}}]{Knutson07b}
{Knutson}, H.~A., {Charbonneau}, D., {Allen}, L.~E., {Fortney}, J.~J., {Agol},
  E., {Cowan}, N.~B., {Showman}, A.~P., {Cooper}, C.~S., \& {Megeath}, S.~T.
  2007{\natexlab{a}}, \nat, 447, 183

\bibitem[{{Knutson} {et~al.}(2009){Knutson}, {Charbonneau}, {Cowan}, {Fortney},
  {Showman}, {Agol}, {Henry}, {Everett}, \& {Allen}}]{Knutson09b}
{Knutson}, H.~A., {Charbonneau}, D., {Cowan}, N.~B., {Fortney}, J.~J.,
  {Showman}, A.~P., {Agol}, E., {Henry}, G.~W., {Everett}, M.~E., \& {Allen},
  L.~E. 2009, \apj, 690, 822

\bibitem[{{Knutson} {et~al.}(2007{\natexlab{b}}){Knutson}, {Charbonneau},
  {Noyes}, {Brown}, \& {Gilliland}}]{Knutson07a}
{Knutson}, H.~A., {Charbonneau}, D., {Noyes}, R.~W., {Brown}, T.~M., \&
  {Gilliland}, R.~L. 2007{\natexlab{b}}, \apj, 655, 564

\bibitem[{{Langland-Shula} {et~al.}(2009){Langland-Shula}, {Vogt},
  {Charbonneau}, {Butler}, \& {Marcy}}]{Langland09}
{Langland-Shula}, L.~E., {Vogt}, S.~S., {Charbonneau}, D., {Butler}, P., \&
  {Marcy}, G. 2009, \apj, 696, 1355

\bibitem[{{Lecavelier des Etangs} {et~al.}(2008{\natexlab{a}}){Lecavelier des
  Etangs}, {Pont}, {Vidal-Madjar}, \& {Sing}}]{Lecavelier08a}
{Lecavelier des Etangs}, A., {Pont}, F., {Vidal-Madjar}, A., \& {Sing}, D.
  2008{\natexlab{a}}, \aap, 481, L83

\bibitem[{{Lecavelier des Etangs} {et~al.}(2008{\natexlab{b}}){Lecavelier des
  Etangs}, {Vidal-Madjar}, {D{\'e}sert}, \& {Sing}}]{Lecavelier08b}
{Lecavelier des Etangs}, A., {Vidal-Madjar}, A., {D{\'e}sert}, J.-M., \&
  {Sing}, D. 2008{\natexlab{b}}, \aap, 485, 865

\bibitem[{{Liang} {et~al.}(2004){Liang}, {Seager}, {Parkinson}, {Lee}, \&
  {Yung}}]{Liang04}
{Liang}, M., {Seager}, S., {Parkinson}, C.~D., {Lee}, A.~Y.-T., \& {Yung},
  Y.~L. 2004, \apjl, 605, L61

\bibitem[{{Lodders}(1999)}]{Lodders99}
{Lodders}, K. 1999, \apj, 519, 793

\bibitem[{{Lodders}(2002)}]{Lodders02b}
---. 2002, \apj, 577, 974

\bibitem[{{Lodders}(2009)}]{Lodders09}
---. 2009, ArXiv:0910.0811, in: Formation and Evolution of Exoplanets, R.
  Barnes (ed.), Wiley, Berlin, in press

\bibitem[{{Lodders} \& {Fegley}(2002)}]{Lodders02}
{Lodders}, K. \& {Fegley}, B. 2002, Icarus, 155, 393

\bibitem[{{Lodders} \& {Fegley}(2006)}]{Lodders06}
---. 2006, {Astrophysics Update 2} (Springer Praxis Books, Berlin: Springer,
  2006)

\bibitem[{{Machalek} {et~al.}(2008){Machalek}, {McCullough}, {Burke},
  {Valenti}, {Burrows}, \& {Hora}}]{Machalek08}
{Machalek}, P., {McCullough}, P.~R., {Burke}, C.~J., {Valenti}, J.~A.,
  {Burrows}, A., \& {Hora}, J.~L. 2008, \apj, 684, 1427

\bibitem[{{Madhusudhan} \& {Seager}(2009)}]{Madhu09}
{Madhusudhan}, N. \& {Seager}, S. 2009, \apj, 707, 24

\bibitem[{{Marley} {et~al.}(2007){Marley}, {Fortney}, {Seager}, \&
  {Barman}}]{Marley07b}
{Marley}, M.~S., {Fortney}, J., {Seager}, S., \& {Barman}, T. 2007, in
  Protostars and Planets V, ed. B.~{Reipurth}, D.~{Jewitt}, \& K.~{Keil},
  733--747

\bibitem[{{Marley} {et~al.}(1999){Marley}, {Gelino}, {Stephens}, {Lunine}, \&
  {Freedman}}]{Marley99}
{Marley}, M.~S., {Gelino}, C., {Stephens}, D., {Lunine}, J.~I., \& {Freedman},
  R. 1999, \apj, 513, 879

\bibitem[{{Marley} \& {McKay}(1999)}]{MM99}
{Marley}, M.~S. \& {McKay}, C.~P. 1999, Icarus, 138, 268

\bibitem[{{Menou} \& {Rauscher}(2009)}]{Menou08}
{Menou}, K. \& {Rauscher}, E. 2009, \apj, 700, 887

\bibitem[{{Miller-Ricci} {et~al.}(2009){Miller-Ricci}, {Seager}, \&
  {Sasselov}}]{MillerRicci09a}
{Miller-Ricci}, E., {Seager}, S., \& {Sasselov}, D. 2009, \apj, 690, 1056

\bibitem[{{Pont} {et~al.}(2008){Pont}, {Knutson}, {Gilliland}, {Moutou}, \&
  {Charbonneau}}]{Pont08}
{Pont}, F., {Knutson}, H., {Gilliland}, R.~L., {Moutou}, C., \& {Charbonneau},
  D. 2008, \mnras, 385, 109

\bibitem[{{Prinn} \& {Barshay}(1977)}]{Prinn77}
{Prinn}, R.~G. \& {Barshay}, S.~S. 1977, Science, 198, 1031

\bibitem[{{Rauscher} \& {Menou}(2009)}]{Rauscher09}
{Rauscher}, E. \& {Menou}, K. 2009, ApJ submitted,arXiv:0907.2692

\bibitem[{{Rauscher} {et~al.}(2007){Rauscher}, {Menou}, {Seager}, {Deming},
  {Cho}, \& {Hansen}}]{Rauscher07b}
{Rauscher}, E., {Menou}, K., {Seager}, S., {Deming}, D., {Cho}, J.~Y.-K., \&
  {Hansen}, B.~M.~S. 2007, \apj, 664, 1199

\bibitem[{{Redfield} {et~al.}(2008){Redfield}, {Endl}, {Cochran}, \&
  {Koesterke}}]{Redfield08}
{Redfield}, S., {Endl}, M., {Cochran}, W.~D., \& {Koesterke}, L. 2008, \apjl,
  673, L87

\bibitem[{{Richardson} {et~al.}(2006){Richardson}, {Harrington}, {Seager}, \&
  {Deming}}]{Richardson06}
{Richardson}, L.~J., {Harrington}, J., {Seager}, S., \& {Deming}, D. 2006,
  \apj, 649, 1043

\bibitem[{{Saumon} {et~al.}(2006){Saumon}, {Marley}, {Cushing}, {Leggett},
  {Roellig}, {Lodders}, \& {Freedman}}]{Saumon06}
{Saumon}, D., {Marley}, M.~S., {Cushing}, M.~C., {Leggett}, S.~K., {Roellig},
  T.~L., {Lodders}, K., \& {Freedman}, R.~S. 2006, \apj, 647, 552

\bibitem[{{Saumon} {et~al.}(2003){Saumon}, {Marley}, {Lodders}, \&
  {Freedman}}]{Saumon03}
{Saumon}, D., {Marley}, M.~S., {Lodders}, K., \& {Freedman}, R.~S. 2003, in
  Brown Dwarfs, Proceedings of IAU Symposium \#211, ed.~E.~Martin, (San
  Francisco: Astronomical Society of the Pacific), 345

\bibitem[{{Seager} {et~al.}(2005){Seager}, {Richardson}, {Hansen}, {Menou},
  {Cho}, \& {Deming}}]{Seager05}
{Seager}, S., {Richardson}, L.~J., {Hansen}, B.~M.~S., {Menou}, K., {Cho},
  J.~Y.-K., \& {Deming}, D. 2005, \apj, 632, 1122

\bibitem[{{Seager} \& {Sasselov}(2000)}]{SS00}
{Seager}, S. \& {Sasselov}, D.~D. 2000, \apj, 537, 916

\bibitem[{{Sharp} \& {Burrows}(2007)}]{Sharp07}
{Sharp}, C.~M. \& {Burrows}, A. 2007, \apjs, 168, 140

\bibitem[{{Showman} {et~al.}(2008){Showman}, {Cooper}, {Fortney}, \&
  {Marley}}]{Showman08}
{Showman}, A.~P., {Cooper}, C.~S., {Fortney}, J.~J., \& {Marley}, M.~S. 2008,
  \apj, 682, 559

\bibitem[{{Showman} {et~al.}(2009){Showman}, {Fortney}, {Lian}, {Marley},
  {Freedman}, {Knutson}, \& {Charbonneau}}]{Showman09}
{Showman}, A.~P., {Fortney}, J.~J., {Lian}, Y., {Marley}, M.~S., {Freedman},
  R.~S., {Knutson}, H.~A., \& {Charbonneau}, D. 2009, \apj, 699, 564

\bibitem[{{Showman} \& {Guillot}(2002)}]{Showman02}
{Showman}, A.~P. \& {Guillot}, T. 2002, \aap, 385, 166

\bibitem[{{Sing} {et~al.}(2009){Sing}, {D{\'e}sert}, {Lecavelier des Etangs},
  {Ballester}, {Vidal-Madjar}, {Parmentier}, {Hebrard}, \& {Henry}}]{Sing09b}
{Sing}, D.~K., {D{\'e}sert}, J.-M., {Lecavelier des Etangs}, A., {Ballester},
  G.~E., {Vidal-Madjar}, A., {Parmentier}, V., {Hebrard}, G., \& {Henry}, G.~W.
  2009, A\&A in press, ArXiv:0907.4991

\bibitem[{{Sing} {et~al.}(2008{\natexlab{a}}){Sing}, {Vidal-Madjar},
  {D{\'e}sert}, {Lecavelier des Etangs}, \& {Ballester}}]{Sing08a}
{Sing}, D.~K., {Vidal-Madjar}, A., {D{\'e}sert}, J.-M., {Lecavelier des
  Etangs}, A., \& {Ballester}, G. 2008{\natexlab{a}}, \apj, 686, 658

\bibitem[{{Sing} {et~al.}(2008{\natexlab{b}}){Sing}, {Vidal-Madjar},
  {Lecavelier des Etangs}, {D{\'e}sert}, {Ballester}, \&
  {Ehrenreich}}]{Sing08b}
{Sing}, D.~K., {Vidal-Madjar}, A., {Lecavelier des Etangs}, A., {D{\'e}sert},
  J.-M., {Ballester}, G., \& {Ehrenreich}, D. 2008{\natexlab{b}}, \apj, 686,
  667

\bibitem[{{Snellen} {et~al.}(2008){Snellen}, {Albrecht}, {de Mooij}, \& {Le
  Poole}}]{Snellen08}
{Snellen}, I.~A.~G., {Albrecht}, S., {de Mooij}, E.~J.~W., \& {Le Poole}, R.~S.
  2008, \aap, 487, 357

\bibitem[{{Spiegel} {et~al.}(2007){Spiegel}, {Haiman}, \& {Gaudi}}]{Spiegel07}
{Spiegel}, D.~S., {Haiman}, Z., \& {Gaudi}, B.~S. 2007, \apj, 669, 1324

\bibitem[{{Spiegel} {et~al.}(2009){Spiegel}, {Silverio}, \&
  {Burrows}}]{Spiegel09}
{Spiegel}, D.~S., {Silverio}, K., \& {Burrows}, A. 2009, \apj, 699, 1487

\bibitem[{{Sudarsky} {et~al.}(2003){Sudarsky}, {Burrows}, \&
  {Hubeny}}]{Sudar03}
{Sudarsky}, D., {Burrows}, A., \& {Hubeny}, I. 2003, \apj, 588, 1121

\bibitem[{{Sudarsky} {et~al.}(2000){Sudarsky}, {Burrows}, \& {Pinto}}]{Sudar00}
{Sudarsky}, D., {Burrows}, A., \& {Pinto}, P. 2000, \apj, 538, 885

\bibitem[{{Swain} {et~al.}(2008){Swain}, {Vasisht}, \& {Tinetti}}]{Swain08}
{Swain}, M.~R., {Vasisht}, G., \& {Tinetti}, G. 2008, \nat, 452, 329

\bibitem[{{Swain} {et~al.}(2009){Swain}, {Vasisht}, {Tinetti}, {Bouwman},
  {Chen}, {Yung}, {Deming}, \& {Deroo}}]{Swain09}
{Swain}, M.~R., {Vasisht}, G., {Tinetti}, G., {Bouwman}, J., {Chen}, P.,
  {Yung}, Y., {Deming}, D., \& {Deroo}, P. 2009, \apjl, 690, L114

\bibitem[{{Tinetti} {et~al.}(2007{\natexlab{a}}){Tinetti}, {Liang},
  {Vidal-Madjar}, {Ehrenreich}, {Lecavelier des Etangs}, \&
  {Yung}}]{Tinetti07a}
{Tinetti}, G., {Liang}, M.-C., {Vidal-Madjar}, A., {Ehrenreich}, D.,
  {Lecavelier des Etangs}, A., \& {Yung}, Y.~L. 2007{\natexlab{a}}, \apjl, 654,
  L99

\bibitem[{{Tinetti} {et~al.}(2007{\natexlab{b}}){Tinetti}, {Vidal-Madjar},
  {Liang}, {Beaulieu}, {Yung}, {Carey}, {Barber}, {Tennyson}, {Ribas},
  {Allard}, {Ballester}, {Sing}, \& {Selsis}}]{Tinetti07}
{Tinetti}, G., {Vidal-Madjar}, A., {Liang}, M.-C., {Beaulieu}, J.-P., {Yung},
  Y., {Carey}, S., {Barber}, R.~J., {Tennyson}, J., {Ribas}, I., {Allard}, N.,
  {Ballester}, G.~E., {Sing}, D.~K., \& {Selsis}, F. 2007{\natexlab{b}}, \nat,
  448, 169

\bibitem[{{Vidal-Madjar} {et~al.}(2004){Vidal-Madjar}, {D{\' e}sert},
  {Lecavelier des Etangs}, {H{\' e}brard}, {Ballester}, {Ehrenreich}, {Ferlet},
  {McConnell}, {Mayor}, \& {Parkinson}}]{Vidal04}
{Vidal-Madjar}, A., {D{\' e}sert}, J.-M., {Lecavelier des Etangs}, A., {H{\'
  e}brard}, G., {Ballester}, G.~E., {Ehrenreich}, D., {Ferlet}, R.,
  {McConnell}, J.~C., {Mayor}, M., \& {Parkinson}, C.~D. 2004, \apjl, 604, L69

\bibitem[{{Vidal-Madjar} {et~al.}(2003){Vidal-Madjar}, {Lecavelier des Etangs},
  {D{\' e}sert}, {Ballester}, {Ferlet}, {H{\' e}brard}, \& {Mayor}}]{Vidal03}
{Vidal-Madjar}, A., {Lecavelier des Etangs}, A., {D{\' e}sert}, J.-M.,
  {Ballester}, G.~E., {Ferlet}, R., {H{\' e}brard}, G., \& {Mayor}, M. 2003,
  \nat, 422, 143

\bibitem[{{Visscher} {et~al.}(2006){Visscher}, {Lodders}, \&
  {Fegley}}]{Visscher06}
{Visscher}, C., {Lodders}, K., \& {Fegley}, B.~J. 2006, \apj, 648, 1181

\bibitem[{{Williams} {et~al.}(2006){Williams}, {Charbonneau}, {Cooper},
  {Showman}, \& {Fortney}}]{Williams06}
{Williams}, P.~K.~G., {Charbonneau}, D., {Cooper}, C.~S., {Showman}, A.~P., \&
  {Fortney}, J.~J. 2006, \apj, 649, 1020

\bibitem[{{Yung} {et~al.}(1988){Yung}, {Drew}, {Pinto}, \& {Friedl}}]{Yung88}
{Yung}, Y.~L., {Drew}, W.~A., {Pinto}, J.~P., \& {Friedl}, R.~R. 1988, Icarus,
  73, 516

\bibitem[{{Zahnle} {et~al.}(2009{\natexlab{a}}){Zahnle}, {Marley}, \&
  {Fortney}}]{Zahnle10}
{Zahnle}, K., {Marley}, M.~S., \& {Fortney}, J.~J. 2009{\natexlab{a}},
  ArXiv:0911.0728, ApJL submitted

\bibitem[{{Zahnle} {et~al.}(2009{\natexlab{b}}){Zahnle}, {Marley}, {Freedman},
  {Lodders}, \& {Fortney}}]{Zahnle09}
{Zahnle}, K., {Marley}, M.~S., {Freedman}, R.~S., {Lodders}, K., \& {Fortney},
  J.~J. 2009{\natexlab{b}}, \apjl, 701, L20

\end{thebibliography}

\end{document}